\documentclass[12pt]{article}
\usepackage{jheppub}
\usepackage{amsfonts,ulem,bbold}   
\usepackage{amssymb,amsmath}

\def\l{\left}
\def\r{\right}
\def\fr{\frac}
\def\la{\label}
\def\d{\partial}
\newcommand{\p}{\bar{P}}  
 
\def\be{\begin{equation}}
\def\ee{\end{equation}}
\def\ba{\begin{eqnarray}}
\def\ea{\end{eqnarray}}

\def\mbf{\mathbf}
\newcommand{\sfrac}[2]{{\textstyle\frac{#1}{#2}}}

\title{More on gapped Goldstones at finite density:
More gapped Goldstones} 

\author[a]{Alberto Nicolis,}
\author[a]{Riccardo Penco,}
\author[b]{Federico Piazza,}
\author[a]{and Rachel A. Rosen}

\affiliation[a]{Physics Department and Institute for Strings, Cosmology, and Astroparticle Physics,\\
  Columbia University, New York, NY 10027, USA}

\affiliation[b]{Paris Center for Cosmological Physics and Laboratoire APC, \\
Universit\'{e} Paris 7, 75205 Paris, France}

\abstract{It was recently argued that certain relativistic theories at finite density can exhibit an unconventional spectrum of Goldstone excitations, with gapped Goldstones whose  gap  is exactly calculable in terms of the symmetry algebra. We confirm this result as well as previous ones concerning gapless Goldstones for non-relativistic systems via a coset construction of the low-energy effective field theory.  Moreover, our analysis  unveils {\it additional} gapped Goldstones, naturally as light as the others, but  this time with a model-dependent gap. Their exact number cannot be inferred solely from the symmetry breaking pattern either, but rather depends on the details of the symmetry breaking mechanism---a statement that we explicitly verify with a number of examples. Along the way we provide what we believe to be a particularly transparent interpretation of the so-called inverse-Higgs constraints for spontaneously broken spacetime symmetries.}

\begin{document}
\maketitle
\flushbottom

\section{Introduction and Summary}

Perhaps counterintuitively, some of the most interesting consequences of symmetries in physics arise when symmetries get broken---spontaneously broken, to be precise. This is because of the Goldstone phenomenon, which identifies the existence of certain low-energy excitations as the only consistent way to realize (non-linearly) the broken symmetries.  These symmetries tightly constrain the dynamics of the Goldstone excitations and, as a consequence, the Goldstone sector is a universal and robust component of all systems with spontaneously broken symmetries.

For spontaneously broken {\it internal} symmetries, the Goldstone theorem predicts exactly gapless excitations. For spontaneously broken {\it spacetime} symmetries,  there is a richer set of possibilities.  In fact, in situations in which time-translations are spontaneously broken---for instance, by a cosmological spacetime---there might not be a conserved energy at all, and excitations cannot even be classified in terms of  their ``gaps.''\footnote{For instance, the mass of cosmological perturbations is not well defined: the  mass parameter formally appearing in their Lagrangian can be  changed by time-dependent field redefinitions, which are consistent with all the symmetries since the background depends explicitly on time.}
%
%

In recent work \cite{Nicolis:2012vf} it was shown that, for relativistic systems at finite density,  Goldstone modes associated with certain {\it internal} symmetries can become gapped.  With the benefit of hindsight, this is not entirely surprising.  A system at finite density for a certain charge $Q$ can be modeled via the effective Hamiltonian
\be \label{Htilde}
H' = H-\mu Q \; ,
\ee
where $H$ is the system's original Hamiltonian, and $\mu$ is the chemical potential. This new Hamiltonian is {\it not} invariant under the symmetries of $H$ that do not commute with $Q$. However, at least for small $\mu$, these can still be thought of as {\it approximate} symmetries of $H'.$  In the case that these approximate symmetries are also {\it spontaneously} broken with a symmetry breaking scale much bigger than $\mu$, the corresponding Goldstones will not be exactly gapless, but will have a small gap proportional to the symmetry breaking parameter $\mu$. A similar  phenomenon happens for pions in QCD, which can be identified with the Goldstone bosons associated with chiral symmetry, which, on top of being spontaneously broken, is also explicitly broken by the quark masses in the QCD Lagrangian. As a consequence, the pions are not exactly massless, but acquire a small mass suppressed by these symmetry breaking parameters. The excitations associated with spontaneously broken approximate symmetries are usually referred to as pseudo-Goldstone bosons.

Remarkably, in contrast to more general pseudo-Goldstone bosons, the gapped Goldstones of the finite density systems analyzed in \cite{Nicolis:2012vf} have mass gaps that are exactly determined by the symmetry algebra.   They are given by $\mu$ times numerical factors that depend only on the symmetry group's structure constants and are thus insensitive to quantum corrections.  This non-perturbative result follows from  thinking of the symmetries that do not commute with $Q$ as being {\it spontaneously} broken by a finite $\mu$, rather than explicitly, in the following sense. The introduction of the modified Hamiltonian \eqref{Htilde} can be viewed as a formal trick to select a particular finite-density state $| \mu \rangle$, i.e., the ground state of $H'$.  However, the generator of time-translations is still the original Hamiltonian $H$.  All symmetries that do not commute with $Q$ are broken by the state of the system, but not by the original Hamiltonian.  That is, they are spontaneously broken. The fact that, according to this viewpoint, they are {\it exact} symmetries of the dynamics allows one to derive exact statements for the associated Goldstones' gaps, via a modified Goldstone theorem \cite{Nicolis:2012vf}.  From now on we will thus avoid referring to these excitations as pseudo-Goldstones---we will simply call them gapped Goldstones.

In this work we use ``coset construction" techniques \cite{Callan:1969sn,Coleman:1969sm,Volkov:1973vd,Ogievetsky:1974ab} to build a generic low-energy effective field theory which recovers these gapped Goldstone modes.  We find perfect agreement with the general theorem of \cite{Nicolis:2012vf}.  Moreover, and perhaps more interestingly, we find that in general there are {\it other} gapped modes, which are not predicted by such a theorem, and whose gaps are not fixed by the symmetry breaking pattern, and yet nevertheless belong in the low-energy effective field theory.  Although these modes are not predicted by a Goldstone theorem, we will refer to them as ``Goldstones" since they non-linearly realize some of the broken symmetries, and, in particular, they reduce to standard, gapless Goldstone bosons when the chemical potential is brought to zero.  From this viewpoint, they are on an equal footing with the fixed-gap Goldstones, only they are more difficult to unveil, and their existence is less universal.  

More specifically, the setup we consider in this work is the same as was considered in \cite{Nicolis:2012vf}: a generic Poincar\'e invariant theory with internal symmetries, in a state that {\it (i)} has finite density for one of the corresponding charges, $Q$, and {\it (ii)} breaks $Q$ as well as some of the other charges.\footnote{Our results can be formally extended to situations in which $Q$ is not broken, but in that  case one can classify excitations directly in terms of their eigevalues for the original Hamiltonian, $H$. With this definition of energy all Goldstones are gapless. We discuss this further in sect.~\ref{prob}.}  Then, since by definition the lowest-energy state at finite density is the ground state of \ref{Htilde} for some $\mu$, and since $Q$ is spontaneously broken by assumption, $H$ is also spontaneously broken, and one cannot characterize excitations in terms of eigenvalues of $H$. The best one can do is use the unbroken combination $H'$ as the definition of energy for the excitations. 

In sections \ref{setup}, \ref{action} and \ref{sec:spectrum} of this work we implement this symmetry breaking pattern directly at the level of the coset construction, and we find four different kinds of Goldstone modes:
\begin{enumerate}
\item {\it Linear gapless}: gapless excitations, with a low-momentum dispersion relation $\omega \propto k \,$.
\item {\it Quadratic gapless}:  gapless excitations, with a low-momentum dispersion relation $\omega \propto k^2/\mu \, $.
\item {\it Fixed gap}: gapped excitations, with a low-momentum gap $\omega \sim \mu$, completely determined by the symmetry breaking pattern.
\item {\it Unfixed gap}: gapped excitations, with a low-momentum gap generically of order $\mu$, but dependent on free parameters, and thus potentially tunable.
\end{enumerate}
\vspace{.3cm}
The first two classes of Goldstones were already identified in the classic paper by Nielsen and Chadha \cite{Nielsen:1975hm}, where a counting rule for them was also derived. This counting rule has been made more powerful over the years, most recently in \cite{Watanabe:2011ec,Watanabe:2012hr,Hidaka:2012ym,Schafer:2001bq}. The third class of Goldstones was identified in \cite{Nicolis:2012vf}, where a counting rule was derived\footnote{In a different context~\cite{Alday:2010ku}, it was argued that fixed gap Goldstones can generically arise when a spontaneously broken symmetry does not commute with the Hamiltonian.}.  Our coset analysis confirms these previous results, except for a possible disagreement with the counting rule of classes 1 and 2 proposed by \cite{Watanabe:2011ec,Watanabe:2012hr}.

Perhaps most importantly, our results identify the fourth class of Goldstones.  In this work we provide a counting rule for these Goldstones, in the form of upper and lower bounds:
\be
\label{bound}
n_2 \le n_4 \le n_2 + n_3 \; ,
\ee
where $n_i$ is the number of Goldstones of class $i$.  Where one lands in this range cannot be inferred purely from symmetry considerations. From the high-energy, microscopic viewpoint,  the uncertainty stems from the freedom one has in choosing, for given symmetry breaking pattern, which representation of the symmetry group the order parameter belongs to, at least for models that are amenable to a semiclassical analysis.  From the low-energy, coset-construction viewpoint, the uncertainty stems from the freedom one has in imposing certain ``inverse Higgs" constraints  \cite{Ivanov:1975zq,Low:2001bw}. 

The current literature on inverse Higgs constraints can be ambiguous as to when inverse Higgs constraints {\it can} be imposed, versus when they {\it should} be imposed.   In this work we put forward an interpretation of the inverse Higgs constraints and a prescription for when to impose them which we believe are more transparent.  In particular, we emphasize their being {\it optional} gauge-fixing conditions.   As is well known, when spacetime symmetries are spontaneously broken, the Goldstone fields {\it can} be a redundant parameterization of the physical excitations.   However, as we demonstrate in the examples considered below, whether they {\it are} is a question that depends of the details of the symmetry breaking mechanism, and not just on the symmetry breaking pattern. To be explicit, at least for weakly coupled linear $\sigma$-models, it depends on which representation breaks the symmetries. In cases in which such redundancies are there, they really correspond to gauge transformations one can perform on the Goldstone fields that do not change the physical fluctuations of the order parameters. One can choose gauges that are compatible with all the global symmetries. Such gauge choices are nothing but the inverse-Higgs constraints.

Both the upper and lower bounds of  \eqref{bound} can be saturated, as we show in a number of examples. However, the upper bound can be saturated for {\it any} symmetry breaking pattern, while saturating the lower bound is not always possible, because some of the necessary inverse Higgs constraints might be incompatible with the unbroken symmetries. So, while the upper bound is universal, the lower bound can be raised for certain symmetry breaking patterns.   As suggested by the upper bound, and as we will make more precise below, the class 4 Goldstones  can be thought of as partners of class 2 and class 3 Goldstones.  The lower bound then suggests that the partners of class 2 Goldstones are always there, while one might be able to remove some or all of the partners  of class 3 Goldstones by imposing inverse Higgs constraints.  We confirm these expectations below, giving specific examples in sections \ref{sec:su(2)} and \ref{sec:so(3)}.

Two qualifications are in order.  First, the coset construction yields the Goldstones' low-energy effective action as a derivative expansion. Since some of our Goldstones are gapped, with a gap of order of the chemical potential $\mu$, we have to make sure that $\mu$ is well below the strong coupling scale of the low-energy effective theory, where the derivative expansion breaks down. As we will discuss at some length below, such a situation corresponds, for instance, to relativistic theories with standard spontaneous symmetry breaking (SSB), in which  the turning-on of a small chemical potential does not trigger a phase transition.  Thus, in these theories, one can have a large symmetry breaking scale $f$ controlling the derivative expansion, and a small $\mu \ll f$ acting as an infrared deformation of the theory and controlling the excitations' gaps.  It is somewhat puzzling that the results of \cite{Nicolis:2012vf} are in this respect much more robust than the coset construction, being completely insensitive to how close $\mu$ is to the strong-coupling scale of the Goldstone low-energy effective theory. It would be interesting to understand whether there exists an improved coset construction that can dispose of the $\mu \ll f$ assumption.

Second, and less importantly, given our high-energy upbringing, we might be tempted to refer to an excitation's zero-momentum gap as its ``mass."  We will refrain from doing so.  Since a finite density state breaks (spontaneously) Poincar\'e invariance, its excitations cannot be classified in terms of their masses---there is no ``invariant mass" to talk about.

\vspace{.3cm}

\noindent
{\it Note added:} While this work was in its final stages, ref.~\cite{Watanabe:2013uya} appeared with some overlapping results. Ref.~\cite{Watanabe:2013uya} extends the results of \cite{Nicolis:2012vf} to intrinsically non-relativistic systems, and avoids dealing explicitly with the spontaneous breaking of spacetime symmetries. 
Notice that in the real world Poincar\'e invariance, if broken, is always broken spontaneously rather than explicitly.
Therefore, although some aspects of ref.~\cite{Watanabe:2013uya} are very general, that analysis formally applies only to cases in which the breaking of spacetime symmetries happens at much higher scales than the breaking of internal symmetries under consideration.  In such cases, the Goldstones associated with spacetime symmetries (e.g., phonons) probably can be  ignored, since their interactions are suppressed by a very high symmetry breaking scale. 

In our work we explicitly keep track of all spacetime symmetries, including the spontaneously broken ones. However, we restrict our analysis to cases in which Lorentz boosts are broken only by the finite density of the ground state, rather than, for instance, by an underlying medium.  As we will see, such a restriction is equivalent to imposing certain inverse Higgs constraints. 

\vspace{.3cm}

\noindent
{\it Note added 2:} After this work was completed, it was brought to our attention that the possible existence of Goldstone modes with unfixed gap was first discussed in ref.~\cite{Kapustin:2012cr}. In that paper, the author discusses the unfixed gap Goldstones that are the partners of the class 2 Goldstones in a context where Lorentz symmetry is explicitly broken. As mentioned above, in this work we point out how unfixed mass Goldstones can also arise as partners of class 3 Goldstones, which were not discussed in Ref.~\cite{Kapustin:2012cr}.
%
\vspace{.5cm}

\noindent
{\it Conventions:}  In this work we adopt a $(-,+,+,+)$ signature for the metric.  All internal indices are raised and lowered arbitrarily.  Einstein summation convention is assumed unless otherwise stated.

\section{Setup}\label{setup}
In this section we present the symmetry breaking pattern which will be the starting point of our coset construction: a generic Poincar\'e invariant theory with internal symmetries, broken and unbroken, in a state that has finite density for one internal charge.  We review certain characteristics of symmetry breaking at finite density.  We discuss the relevance of inverse Higgs constraints for this symmetry breaking pattern.  We also discuss an interpretation of the inverse Higgs constraints as a gauge fixing condition.

\subsection{Symmetry breaking pattern} \label{sec:sbp}
As we briefly reviewed in the Introduction, a system at finite density for a conserved charge $Q$ can be described using the modified Hamiltonian
\be \la{Hreplace}
H \rightarrow H' \equiv H - \mu Q \, ,
\ee
where $\mu$ is the chemical potential associated with the charge $Q$. The ground state $| \mu \rangle$ of the system is defined as the eigenstate of the modified Hamiltonian $H'$ with the lowest eigenvalue which, without loss of generality, we can assume to vanish:\footnote{Such a choice can be implemented by dialing the cosmological constant \cite{Nicolis:2011pv}, which, in the absence of gravity, has no physical consequences.}
\be
H' | \mu \rangle = (H-\mu Q )| \mu \rangle = 0\,.
\ee
If $Q$ is spontaneously broken then $H$, the generator of time translations, must be as well, in the sense that $| \mu \rangle$ is not an eigenstate of $H$ {\cite{Nicolis:2011pv}.  Thus excited states, including the Goldstone bosons, cannot be classified as eigenstates of the original Hamiltonian $H$ but only as eigenstates of $H-\mu Q$.

In the Lagrangian formulation, the replacement (\ref{Hreplace}) is equivalent to the following shift of the time derivatives:
\be
\partial_0 \rightarrow \partial_0 + i \mu Q \, .
\ee
Introducing a chemical potential in this way explicitly breaks the Lorentz invariance of the Lagrangian.  However, when $Q$ is spontaneously broken, a completely equivalent description is one in which the Lagrangian is the original, Lorentz-invariant one, but one expands about a {\it time-dependent} background solution,
\be \la{SSPdef}
\langle \Phi \rangle (t) =e^{i\mu Q t} \langle \Phi \rangle_0 \, , 
\ee
where $\Phi$ is the order parameter of the symmetry breaking and $\langle \Phi \rangle_0$ is its expectation value at $t=0$.  A field configuration of the form (\ref{SSPdef}) was dubbed ``spontaneous symmetry probing" (SSP) in {\cite{Nicolis:2011pv}.  In this approach, one may consider the Lorentz invariance of the theory to be spontaneously broken by a time-dependent field configuration.  

Notice that both the Lagrangian and Hamiltonian formulations allow for an interpretation of our physical situation in which all symmetries that are broken are only broken spontaneously.  According to this interpretation, the original time translations are also spontaneously broken. There is, however, a new generator of time translations that is unbroken: $H'$.  This is the interpretation that we will take in the rest of this paper.

To study gapped Goldstones then, we consider a generic Poincar\'e invariant theory with internal symmetries, in a state that has finite density for one of the internal charges, $Q$.  We take the ground state of the system to spontaneously break $Q$, time translations and boosts, as well as some additional internal charges.  

To implement the coset construction for such a symmetry breaking pattern, it is helpful to pick a convenient basis for the generators of the internal symmetries.  Let us denote the full symmetry group of the Lagrangian by $G$ with generators $Q_I$, and the group of internal symmetries that are left unbroken by the ground state $| \mu \rangle $ by $G'$ with generators $T_A$. We can always choose the unbroken generators in such a way that the structure constants of $G'$ are totally antisymmetric, and then choose the broken generators in such a way that the remaining structure constants of the full group $G$ are also totally antisymmetric.\footnote{As usual, we are implicitly assuming that these internal symmetry groups are products of simple compact Lie groups ($SU(n)$, $SO(n)$, etc.) and $U(1)$ factors.} Hence, in what follows we will assume that all structure constants are totally antisymmetric. 

In general, the charge $Q$ at finite density is given by the sum of one broken (internal) generator $X$ and one unbroken (internal) generator $T$:\footnote{Broken generators are defined up to a combination of unbroken ones, so one could chose a basis of generators containing directly $Q$, and effectively set $\mu_T = 0$.  However, there are other requirements on the basis of generators, that we find more important: $a)$ the basis should include the maximal number of unbroken generators, this is needed for the coset construction; $b)$ the structure constant of the Lie algebra should be totally anti-symmetric in the chosen basis, this facilitates the calculations. In general, it is not always possible to satisfy $a)$ and $b)$ together with $\mu_T = 0$.}
\be \la{Q}
\mu Q =  \mu_X X + \mu_T T.
\ee
This leads us to consider the following pattern of symmetry breaking as the starting point of our construction:
\ba \la{pattern}
\begin{array}{lcl}
\mbox{unbroken} &=&  \l\{
\begin{array}{ll}
\bar{P}^0 \equiv H - \mu_X X - \mu_T T  &  \qquad \mbox{time translations} \\
\bar{P}^i \equiv P^i &  \qquad \mbox{spatial translations}  \\
J_i &  \qquad \mbox{rotations}\\
T_A &  \qquad \mbox{internal symmetries (including $T$)}
\end{array}
\r. 
\\ && \\
\mbox{broken} &=&  \l\{
\begin{array}{ll}
K_i &  \qquad\qquad\qquad\qquad\quad\,\,\,\,   \mbox{boosts} \\
X,X_\textsf{a} &  \qquad\qquad\qquad\qquad\quad \,\,\,\,  \mbox{internal symmetries}
\end{array}
\r.
\\ && \\
\end{array}
\ea
where we denoted with $X_\textsf{a}$ all the broken internal generators but $X$, since in what follows the latter will play a special role.

As mentioned above, we take the boost symmetry to be spontaneously broken by the finite charge density, rather than explicitly broken. Therefore, following the standard procedure~\cite{Ogievetsky:1974ab,Volkov:1973vd}, we parametrize the coset $G/G'$ as:
\be \la{cospar}
\Omega = e^{i x^\mu \p_\mu} e^{i \pi(x) X} e^{i \pi^\textsf{a}(x) X_\textsf{a}} e^{i \eta^i(x) K_i} \, ,
\ee
where $\pi$, $\pi^\textsf{a}$, and $\eta^i$ are the Goldstone fields.

\subsection{Inverse Higgs constraints as gauge choices} \label{IH}

Not all of the Goldstone fields that appear in the coset parametrization (\ref{cospar}) necessarily correspond to independent propagating degrees of freedom. This is due to the well known fact that, whenever spacetime symmetries are spontaneously broken, there can be fewer Goldstone modes than broken generators \cite{Volkov:1973vd,Nielsen:1975hm,Ivanov:1975zq,Low:2001bw}.  A classic example of this phenomenon is provided by a $(d-1)$-brane in $(d+1)$ spacetime dimensions.   Despite breaking $(d+1)$ spacetime generators (one translation, one boost, $(d-1)$ rotations), this system can be described at low energies by only one Goldstone field: the brane's position in the transverse direction \cite{Low:2001bw}.

At least from a semiclassical viewpoint, this mismatch between the number of Goldstone modes and broken generators can happen because linearly independent broken generators need not generate linearly independent local fluctuations when acting on a coordinate-dependent expectation value of the order parameter~\cite{Low:2001bw}. In our particular case, this means that the equation
\be \la{fluctuations}
0 = \delta \Phi(x) \approx \big( \pi(x) X + \pi^\textsf{a} (x )X_\textsf{a} + \eta^i (x) K_i \big) \langle \Phi \rangle (t) \; ,
\ee
where $\langle \Phi \rangle$ is defined in equation (\ref{SSPdef}), can have some non-trivial solutions, i.e.~solutions with non-vanishing Goldstones fields. 

For the symmetry breaking pattern we consider in this work (\ref{pattern}), one such solution follows immediately from the fact that our $\langle \Phi \rangle$ depends only on time.  Setting $\pi^\textsf{a}=0$ and using that, for spacetime scalar functions, $K_i = i \big(t \partial_i -  x^i \partial_t )$, $X = (H-H'-\mu_T T)/\mu_X$, $H = i \partial_t$, and that $H'$ and $T$ are unbroken, equation (\ref{fluctuations}) becomes
\be \label{pi and eta}
\l( \fr{\pi}{\mu_X} - \eta^i x_i \r)   \langle \dot \Phi \rangle = 0 \; .
\ee
This clearly admits the non-trivial solution $\pi = \mu_X \, \eta^i \, x_i$, for any $\eta^i(x)$. The meaning of this solution is that a localized fluctuation of $\langle \Phi \rangle$ parameterized by  arbritrary  $\eta^i$ fields and vanishing $\pi$, can be parameterized equally well by vanishing $\eta^i$ and non-vanishing $\pi$, with $\pi_{\rm new} = - \mu_X \, \eta_{\rm old}^i \, x_i$. Therefore, the fields $\eta^i(x)$ do not describe physically independent  fluctuations and, equivalently, the spectrum of low-energy excitations does not contain independent Goldstone particles associated with the breaking of boosts. We would like to emphasize that even this conclusion follows from some implicit assumptions, for instance, that our order parameter $\Phi$ is a spacetime scalar.  This is not necessary since boosts are spontaneously broken.  In principle, there can be other consistent scenarios in which the $\eta^i $ describe physically independent excitations \cite{framids}.

In general, some of the $\pi^\textsf{a}$ may also describe redundant fluctuations of the order parameter. However, whether or not that is the case depends not only on the pattern of symmetry breaking, but also on the representation of the internal symmetry group furnished by the order parameter. In other words, perhaps not surprisingly, the number of non-trivial solutions to equation (\ref{fluctuations}) depends in general on which representation  $\Phi$ belongs to, in line with our comment above about the importance of $\Phi$ being a scalar. In section \ref{sec:so(3)} we will illustrate this point with an explicit example.

At the level of the coset construction, the ``unphysical'' Goldstone modes are eliminated from the effective action by setting to zero the covariant derivatives of some of the Goldstone modes in a way that is manifestly invariant under the unbroken group. The conditions obtained this way are known as ``inverse Higgs constraints" and are also invariant under all the non-linearly realized symmetries~\cite{Ivanov:1975zq}. Operationally, anytime the commutator between an unbroken momentum $\bar{P}$ and a broken generator $X$ yields another broken generator $X'$, i.e.
\be \label{commute}
[\bar{P}, X ] \sim X' + \dots,
\ee
and $X$ and $X'$ do not belong to the same irreducible multiplet under the unbroken symmetries, one {\it can} impose an inverse Higgs constraint of the form $D \pi'=0$, where $D$ is a covariant derivative operator.  In this way one can express $\pi$ in terms of derivatives of $\pi'$. So, for instance, for our symmetry breaking pattern (\ref{pattern}) we have 
\be
[ P_i , K_j ] = - i \delta_{ij} (\bar{P}_t  -\mu_X X - \mu_T T),
\ee
which means that one can express the Goldstone fields $\eta^i$ associated with the boosts in terms of derivatives of $\pi$ by solving the constraint $D_i \pi = 0$.

However, there is a fair degree of uncertainty in the literature on whether, for a given symmetry breaking pattern, the possible inverse Higgs constraints are something that one {\it (i)} should always impose, {\it (ii)} can impose at will, but can also choose not to impose, or {\it (iii)} can, at times, ignore because, even when not imposed, they may arise automatically from the unconstrained equations of motion \cite{McArthur:2010zm}. The inverse Higgs ideology itself is confusing.  It is usually phrased as the statement that, since covariant derivatives transform covariantly under all the symmetries, unbroken and broken alike, it is {\it consistent} with the symmetries to set some of them to zero.  But why should we start setting things to zero by hand in the first place? In theories {\it without} symmetries, where we are not constrained to make choices that are consistent with the symmetries, we don't simply set to zero arbitrary combinations of fields and derivatives. Why should we start now?

We feel that the following considerations provide a more lucid assessment of the situation.
If we go back to the example of the boost Goldstone fields analyzed above, we see that we can rephrase the existence of non-trivial solutions to eq.~\eqref{fluctuations} as a statement of {\it gauge redundancy}: the physical fluctuation $\delta \Phi$ is invariant under the simultaneous replacement
\be
\vec \eta(x) \to \vec \eta(x) + \vec \epsilon \, (x) \; , \qquad \pi(x) \to \pi(x) + \mu_X \, \vec x \cdot \vec \epsilon \, (x) \; ,
\ee
where $\vec \epsilon$ is a generic (vector) function of $t$ and $\vec x$. This tells us that, in this example, the $\eta^i$ are redundant fields because their spacetime dependence can be changed at will by a suitable choice of $\vec \epsilon$. 
In other words, they are pure gauge fields. 

In the next section we will see that the associated inverse Higgs constraint, to linear order, takes the form
\be
D_i \pi = \partial _i \pi - \mu_X \, \eta^i + \dots = 0  \; .
\ee
This ``constraint" should be thought of as a {\it gauge choice}. In particular, being defined in terms of the vanishing of a covariant derivative, it is a gauge choice that is consistent with all the (global) symmetries.
For more general systems such that the $\eta^i$ are {\it not} pure gauge fields \cite{framids}, there is no gauge redundancy to begin with, in the sense that $\pi$ and $\eta^i$ parameterize truly independent physical fluctuations $\delta \Phi$.  Thus there is no need to fix any gauge via an inverse Higgs constraint. 

Identical considerations apply to the $\pi^\textsf{a}$ fields. As we will see in sect.~\ref{sec:so(3)}, for a given symmetry breaking pattern, there are some systems in which some of the $\pi^\textsf{a}$ are pure gauge, and some systems in which all of them are physical.

It should  now be clear that, at the level of the coset construction, if one  is only given the symmetry breaking pattern and no further information on how the symmetries are broken---for instance, which representations are involved---one has to entertain the possibility that there are gauge redundancies that make certain Goldstones pure gauge. Whether this possibility is there, and which Goldstones it involves, is signaled by which gauge choices are consistent with the global symmetries, i.e.,~by the set of allowed inverse Higgs constraints, which is  determined by the procedure outlined above, in the paragraphs immediately before and after \eqref{commute}.  By {\it choosing} to impose some or all of the allowed inverse Higgs constraints, one is effectively doing two  things: {\it (i)} declaring that, yes, there are gauge redundancies and certain Goldstones are pure gauge fields, and {\it (ii)} choosing gauge-fixing conditions that remove these redundancies in a way that is consistent with all the global symmetries. Notice that, unlike gauge redundancies involving relativistic spin-one gauge fields, which cannot be completely gauge-fixed directly at the level of the action without giving up locality or Lorentz-invariance, our gauge redundancies are similar to that of a St\"uckelberg scalar in a massive gauge theory, for which one {\it can} consistently choose the unitary gauge directly at the level of the action.

Not imposing any of the inverse Higgs constraints is also a consistent choice, which in general defines a {\it physically different} system, because now more physical degrees of freedom are involved. Once again, this will be manifest in the examples of sect.~\ref{sec:so(3)}.  This is an important point, because it means that, unlike in the case of purely internal symmetries~\cite{Coleman:1969sm,Callan:1969sn}, spacetime symmetries potentially admit several inequivalent non-linear realizations.  One can have a number of Goldstone fields each realizing non-linearly  {\it several} broken symmetries, or the usual one-to-one correspondence between Goldstone fields and broken generators.

Given that an inverse Higgs constraint is a gauge-fixing condition or a gauge choice, rather than a ``constraint" in the usual sense, and given that the ``inverse Higgs" part of its name is also potentially unclear, 
in the following we will refer to the imposing of an inverse Higgs constraint as ``fixing (or choosing) a gauge."  
%

\section{Low-Energy Effective Action}\label{action}
In this section we construct the generic low-energy effective action that realizes the symmetry breaking pattern (\ref{pattern}).  To do so, we adopt the coset construction of Callan, Coleman, Wess, and Zumino \cite{Coleman:1969sm,Callan:1969sn} for spacetime symmetries~\cite{Ogievetsky:1974ab,Volkov:1973vd}.  We discuss the appropriate choices of coefficients, technical naturalness and the strong coupling scale of this effective action.

\subsection{The coset construction}
In order to construct an effective action that is invariant under the full symmetry group $G$, one considers the Maurer-Cartan form $\Omega^{-1} d\Omega$, where $\Omega$ is the parametrization of the coset given in (\ref{cospar}).  The Maurer-Cartan form is then expanded in the basis of unbroken and broken generators:
\be
\Omega^{-1} \partial_\nu \Omega = i e_\nu^{~\mu} \l( \p_\mu + A^i_\mu J_i +B^A_\mu T_A  + \,  D_\mu \pi \, X +  D_\mu \pi^\textsf{a}  X_\textsf{a} + D_\mu \eta^i   K_i  \r) \, .
\ee
The $e_\nu^{~\mu}$ are spacetime vierbeins.  The coefficients of the broken generators $D_\mu \pi, D_\mu \pi^\textsf{a}$ and $D_\mu \eta^i$  are the covariant derivatives of the corresponding Goldstone fields $\pi$, $\pi^\textsf{a}$, and $\eta^i$ respectively. These covariant derivatives transform covariantly under {\it all} the symmetries, including the spontaneously broken ones, and can thus be used as the building blocks of the invariant Lagrangian. 

In particular, if we combine these covariant derivatives into structures that are manifestly invariant under the {\it unbroken} symmetries only, then they will automatically be invariant under the broken ones as well. From this viewpoint, it is somewhat misleading that we are still using a relativistic notation for the spacetime index $\mu$: since Lorentz invariance is spontaneously broken, the $\mu=0$ and $\mu=i$ components of the covariant derivatives have to be treated as independent. Whatever combination we write down that is invariant under the unbroken rotations will also be invariant under Lorentz boosts.

In order to explicitly calculate the covariant derivatives for the Goldstones, we will need as much information as possible about the structure constants of the internal symmetry group. As mentioned earlier, we have chosen the internal generators in such a way that the structure constants are totally antisymmetric.  Now, since the $T_A$ span the subgroup $G'$, their algebra does not involve the broken generators $X$ and $X_\textsf{a}$:
\be \la{Talgebra}
[ T_A, T_B ] = i f_{ABC} T^C.
\ee
The finite density state $| \mu \rangle$ must be a simultaneous eigenstate of  $H - \mu Q$ and all $T_A$, because such charges are unbroken. Therefore, we get:
\be \la{comm}
0 = [H - \mu_X X - \mu_T T, T_A ] | \mu \rangle = - i \mu_X f_{X A \textsf{b}} X^\textsf{b} | \mu \rangle \, ,
\ee
where the index $X$ in $f_{X A \textsf{b}}$ is associated with the generator $X$.  If some of the structure constants $ f_{X A \textsf{b}}$ were nonzero, then equation (\ref{comm}) would imply the existence of some linear combinations of the broken generators $X^\textsf{b}$ that remain unbroken. However, by construction the $T_A$ are the maximum number of linearly independent unbroken generators, and therefore we must have $f_{X A \textsf{b}} = 0$. By combining this result with equation (\ref{Talgebra}) and using the total antisymmetry of the structure constants, we conclude that 
\be
[T_A, X ] = 0. 
\ee
In particular, this means that $[T,X] =0$. Finally, the total antisymmetry of the structure constants also implies that the generators $X_\textsf{a}$ must transform according to a (possibly reducible) representation of the unbroken group $G'$:
\be
[ T_A, X_\textsf{a} ] = i f_{A\textsf{a}\textsf{b}} X^\textsf{b}.
\ee
We are now in a position to calculate the Maurer-Cartan form: 
\ba \label{MC}
\Omega^{-1}\partial_\mu\Omega 
&=& i \Lambda_\mu {}^{\nu}\p_\nu - i\mu_X\Lambda_\mu {}^{0} X - i\mu_T\Lambda_\mu {}^{0} T + i \mu_T \delta_\mu^0 \Omega_X^{-1} T \Omega_X 
  \nonumber\\
&& + \Omega_X^{-1}\partial_\mu \Omega_X + i \left(\partial_\mu\pi + \mu_X \delta_\mu^{0}\right)\Omega_X^{-1}X\Omega_X+ \Omega_K^{-1}\partial_\mu \Omega_K \, ,
\ea
where $\Omega_X \equiv e^{i \pi_\textsf{a} X_\textsf{a}}$, $\Omega_K \equiv e^{i\eta^iK_i}$, and we used that $\Omega_K^{-1} P_\mu \Omega_K =\Lambda_\mu {}^{\nu} (\eta) P_\nu$, with
\begin{subequations} \label{Lam}
\begin{align}
&\Lambda_0{}^0 = \cosh \eta, \qquad \Lambda_0{}^i = \eta^i \, \sfrac{\sinh \eta}{\eta} & \\
&  \Lambda_i{}^0 = \eta_i \, \sfrac{\sinh \eta}{\eta}, \qquad \Lambda_i{}^j = \delta_i^j - \eta_i \eta^j \, \sfrac{1- \cosh \eta}{\eta^2}   \, . &
\end{align}
\end{subequations}
where $\eta \equiv \sqrt{\vec \eta \, ^2}$. Notice that all functions of $\eta$ appearing above are even in $\eta$, and thus analytic in $\vec \eta$.

Let us focus on the covariant derivatives for the $\pi$ and $\pi^\textsf{a}$ fields, i.e., the coefficients of the generators $X$ and $X_\textsf{a}$ respectively.  Since our ultimate  goal is to obtain the dispersion relations for these fields, we need only determine their covariant derivatives up to second order in the fields.  Accordingly, we expand the above objects to second order:
\be \label{Om}
\begin{array}{lcl}
\Omega_X^{-1}X\Omega_X & \simeq & X-f_{X\textsf{a}I}\pi^\textsf{a}Q_I
 +\tfrac{1}{2}f_{X\textsf{a}I}f_{I\textsf{b}J}\pi^\textsf{a}\pi^\textsf{b}Q_J \, ,\\
\Omega_X^{-1}T\Omega_X & \simeq &T-f_{T\textsf{a}I}\pi_\textsf{a}Q_I
 +\tfrac{1}{2}f_{T\textsf{a}I}f_{I\textsf{b}J}\pi_\textsf{a}\pi_\textsf{b}Q_J\, , \\
\Omega_X^{-1}\partial_\mu \Omega_X & \simeq & i(\partial_\mu\pi_\textsf{a} X_\textsf{a}
-\tfrac{1}{2}f_{\textsf{a}\textsf{b}I}\pi_\textsf{b} \partial_\mu \pi_\textsf{a} Q_I) \, .\\
\end{array}
\ee
Since the nested commutators of the $K_i$'s that we would get from expanding $ \Omega_K^{-1}\partial_\mu \Omega_K$ only yield $K$'s and $J$'s, that part of the Maurer-Cartan form  will not contribute to the covariant derivatives of the $\pi$ and $\pi^\textsf{a}$ fields, which are our primary interest in this work.

Before using these expressions to determine  $D_\mu \pi$ and  $D_\mu \pi^\textsf{a}$, there is one subtlety we must address.  As briefly reviewed in sect.~\ref{IH}, a consequence of having broken spacetime symmetries is that the Goldstone fields $\eta^i$ associated with the broken boosts are not independent degrees of freedom.   We can eliminate them in favor of the ``physical" Goldstone field $\pi$ by choosing a gauge fixing condition of the form $D_i \pi = 0$. To linear order, we can solve this constraint for $\eta^i$ and get
\be \label{IHeta}
D_i \pi=0 ~~~ \Rightarrow ~~~ \eta_i \simeq \frac{\partial_j \pi}{\mu_X} \, .
\ee
This shows that the covariant derivatives $D_\mu \eta_i$ contain more than one derivative per field, and therefore can be neglected compared to the covariant derivatives of $\pi$ and $\pi^\textsf{a}$ at sufficiently low energies.

Note that if $\mu_X=0$ the replacement (\ref{IHeta}) would not be possible, but then again, for $\mu=0$ boosts are unbroken---at least according to our assumptions---and the associated Goldstone fields are not there in the first place. We will comment further about this discontinuity for $\mu \to 0$ in sect.~\ref{sec:scs}.

By using the results (\ref{Om}), (\ref{Lam}), and (\ref{IHeta}) in expression (\ref{MC}), we can read off the remaining covariant derivatives for the $\pi$ and $\pi^\textsf{a}$:
\be
\begin{array}{ccl} \label{covdev1}
D_0 \pi & \simeq & \dot{\pi}-\frac{1}{2\mu_X}(\partial_j \pi)^2
-\frac{1}{2}f_{X\textsf{a}\textsf{b}}\dot{\pi}_\textsf{a} \pi_\textsf{b}
- \frac{1}{2}( \mu_X f_{X\textsf{a}I}+\mu_Tf_{T\textsf{a}I})f_{X\textsf{b}I}\pi_\textsf{a} \pi_\textsf{b}  \, ,  \vspace{.2cm}\\
D_0 \pi_\textsf{a} &\simeq &  \dot{\pi}_\textsf{a} - \frac{1}{\mu_X}(\partial_j \pi)(\partial_j \pi_\textsf{a}) 
+(\mu_X f_{X\textsf{a}\textsf{b}}+\mu_T f_{T\textsf{a}\textsf{b}})\pi_\textsf{b}
+ f_{X\textsf{a}\textsf{b}}\dot{\pi}\pi_\textsf{b}\\
&&-\frac{1}{2}f_{\textsf{a}\textsf{b}\textsf{c}}\dot{\pi}_\textsf{b} \pi_\textsf{c}
-\frac{1}{2}( \mu_X f_{X\textsf{b}I}+\mu_Tf_{T\textsf{b}I})f_{\textsf{a}\textsf{c}I}\pi_\textsf{b} \pi_\textsf{c}  \, ,  \vspace{.2cm} \\
D_j \pi_\textsf{a} & \simeq & \partial_j \pi_\textsf{a}  \, . 
\end{array}
\ee

The combination of structure constants $\mu_X f_{X \textsf{a}\textsf{b}} + \mu_T f_{T \textsf{a}\textsf{b}}$ appears repeatedly in the above covariant derivatives and it arises from the commutator between $Q$ and the broken charges $X_\textsf{a}$:
\be
\mu [Q, X_\textsf{a}] = i (\mu_X f_{X \textsf{a}\textsf{b}} + \mu_T f_{T \textsf{a}\textsf{b}}) X^\textsf{b}.
\ee
We can then simplify the covariant derivatives by rotating the broken generators $X_\textsf{a}$ in such a way that the matrix $ M_{\textsf{a}\textsf{b}}\equiv  \mu_X f_{X \textsf{a}\textsf{b}} + \mu_T f_{T \textsf{a}\textsf{b}}$ becomes block diagonal:
\be \la{block}
 M_{\textsf{a}\textsf{b}}=  \mu \, \mbox{diag} \l\{ 0, \cdots, 0, 
\l(
\begin{array}{cc}
0 &q_1\\
- q_1 & 0 
\end{array}
\r), \cdots, 
\l(
\begin{array}{cc}
0 &q_k\\
- q_k & 0 
\end{array}
\r)\r\}.
\ee
This can always be achieved because of the antisymmetry of  $ F_{\textsf{a}\textsf{b}}$.
Moreover, we can assume without loss of generality that all $q_n$'s are positive. Equation (\ref{block}) then suggests that we should split the Goldstone bosons $\pi^\textsf{a}$ into two groups, depending on whether the associated generators commute with $Q$ or not. This split is very useful because it allows us to derive additional constraints on the structure constants that follow from equation (\ref{block}) and total antisymmetry and that would not be apparent otherwise. Let us denote the modes corresponding to commuting generators by  $\pi^\alpha$ and the modes corresponding to non-commuting generators by $\pi^a_\pm$. Notice that the non-commuting modes always come in pairs $(\pi^a_+, \pi^a_-)$, and each pair corresponds to a different block on the RHS of equation (\ref{block}). Then, by using the Jacobi identity satisfied by the structure constants, one can show that
\begin{subequations}
\ba
f_{\beta\gamma a_\pm} &=& f_{X\gamma a_\pm} = 0 \la{f1}\\
f_{\alpha a_+ b_+} &=& f_{\alpha a_- b_-} \propto \delta_{q_a  q_b} \\
f_{\alpha a_- b_+} &=& f_{\alpha b_- a_+} \propto \delta_{q_a  q_b} \\
f_{X a_+ b_+} &=& f_{X a_- b_-} \propto \delta_{q_a  q_b} \\
f_{X a_- b_+} &=& f_{X b_- a_+} \propto \delta_{q_a  q_b} \; .
\ea
\end{subequations}
A detailed derivation of these results is provided in Appendix \ref{app1}. The covariant derivatives (\ref{covdev1}) can now be rewritten in terms of the fields $\pi, \pi^\alpha, \pi_\pm^a$ as follows:
\begin{subequations} \la{covdev}
\ba
D_0 \pi &\simeq& \dot{\pi} - \frac{1}{2\mu_X}\partial_j \pi \partial^j \pi -\frac{1}{2}f_{X\alpha\beta}\dot{\pi}^\alpha \pi^\beta  - \frac{1}{2}f_{Xa_+b_-}\left[(D_0 \pi^a_+)\pi^b_- - (D_0 \pi^a_- )\pi^b_+   \right]\qquad \nonumber  \\
&& -\frac{1}{2}f_{X a_+b_+} \left[(D_0 \pi^a_+)\pi^b_+ + (D_0 \pi^a_- )\pi^b_- \right] \la{cd1} \\ 
D_0 \pi_\alpha &\simeq& \dot{\pi}_\alpha-\frac{1}{\mu_X}\partial_j \pi \partial^j \pi_\alpha -\frac{1}{2} f_{\alpha\beta\gamma}\dot{\pi}^\beta \pi^\gamma
-\frac{1}{2}f_{\alpha a_+ b_-}\left[(D_0 \pi^a_+)\pi^b_- - (D_0 \pi^a_- )\pi^b_+   \right] \qquad \nonumber  \\
&& -\frac{1}{2}f_{\alpha a_+b_+} \left[(D_0 \pi^a_+)\pi^b_+ + (D_0 \pi^a_- )\pi^b_- \right]  + f_{X\alpha\beta} \dot{\pi} \pi^\beta\, , \la{cd2}\\
D_0 \pi_a^+ &\simeq&  \dot{\pi}_a^+ + \mu q_a\pi_a^- \, ,\\
D_0 \pi_a^- &\simeq&  \dot{\pi}_a^- - \mu q_a\pi_a^+ \, ,\\
D_j \pi_\alpha & \simeq &\partial_j \pi_\alpha \, , \la{cd3} \\
D_j \pi_a^+ &\simeq& \partial_j \pi_a^+  \, , \\
D_j \pi_a^- &\simeq&  \partial_j \pi_a^- \, .
\ea
\end{subequations}
Since in the next section we will be interested in studying the spectrum of Goldstone modes, we are keeping only the terms that can contribute to the quadratic Lagrangian. In particular, for the $D_j$ covariant derivatives, we only need to keep the terms up to first order in the Goldstones, because the $D_j$'s always have to appear in pairs, to preserve the unbroken rotational invariance. Likewise, the $D_0 \pi_a^\pm$ derivatives contain linear terms without derivatives, i.e., potential tadpole terms, which we can avoid only if we multiply $D_0 \pi_a^\pm$ by another covariant derivative, or by itself.

Based on our experience with the chiral Lagrangian and other effective theories for Goldstone bosons, we may be tempted to conclude that the low energy effective action for the Goldstones should have the schematic form
\be \la{naiveguess}
S \stackrel{?}{=} f^4 \int d^4 x \, \mathcal{L} (D \pi /f),
\ee
where $D\pi$ stands for any of the covariant derivatives in (\ref{covdev}), which of course must be contracted in such a way that the action be explicitly invariant under the unbroken symmetries. The scale $f$ is some symmetry breaking scale which, loosely speaking, can be thought of as the ``size'' of the order parameter and is the analog of the pion's decay constant. As we mentioned in the introduction, this scale does not need to coincide with $\mu$, which is instead the  scale associated with the time-dependence of the order parameter, as shown in equation (\ref{SSPdef}), and thus with the breaking of boosts. However, when $f \gg \mu$, the naive guess (\ref{naiveguess}) gives rise to superluminal modes unless some of the coefficients in the Lagrangian are tuned to be of order $\mu /f \ll 1$.

To illustrate this point, let us focus on the mode $\pi$ and neglect the mixing with other modes. For simplicity, we will also assume that $\mu \sim \mu_X \sim \mu_T$.  The action (\ref{naiveguess}) contains then the following terms quadratic in $\pi$:
\ba \nonumber
S \ \supset \ f^4 \int \l[ c_1 \fr{D_0 \pi}{f} + c_2 \fr{(D_0 \pi)^2}{f^2}  \r] \ \supset \ f^2 \int \l[ c_2 \dot{\pi}^2 -  \fr{c_1 f}{2 \mu_X} (\d_j \pi)^2 \r].
\ea
Clearly, the propagation speed  $c_\pi^2 \sim   \fr{c_1 f}{c_2 \mu}$ can be subluminal only if 
$
c_1/c_2  \sim \mu/f \ \ll\ 1.
$
The reason why this tuning is not only necessary, but also technically natural, is that it can be protected by the spurionic ``CT" symmetry
\be \la{Z}
\mu \to - \mu \; ,\qquad \qquad \qquad t \to - t \; .
\ee
As can be seen from equations (\ref{covdev}), all the time components of the covariant derivatives are odd under the symmetry transformation (\ref{Z}). This means that any term in the Lagrangian  containing an odd number of covariant time derivatives must come with an odd number of factors of $\mu/f$ if the Lagrangian is to be invariant under (\ref{Z}). More precisely, if the terms linear in $D_0 \pi$ and $D_0 \pi^\alpha$ are suppressed by one power of $\mu/f$ compared to the quadratic ones, it is easy to convince oneself that quantum corrections will generate all the other odd terms with a $\mu/f$-suppressed coefficient.

In conclusion, the most general action that we can write down using the covariant derivatives (\ref{cd1}), (\ref{cd2}), (\ref{cd3}) and (\ref{cd4}), that does not contain any tadpole term, is manifestly invariant under the unbroken symmetries, and contains at most two derivatives,~is
\ba
S  &=& f^2 \int d^4 x \bigg\{ \mu_X \,  b  D_0 \pi +  \mu_X \, b_\alpha D_0 \pi^\alpha + c \, (D_0 \pi)^2 + c_\alpha D_0 \pi^\alpha D_0 \pi + c_{\alpha\beta} D_0 \pi^\alpha D_0 \pi^\beta \nonumber\\
   && + c_a D_0 \pi^a  D_0 \pi  + c_{a \alpha} D_0 \pi^a D_0 \pi^\alpha  +c_{ab} D_0 \pi^a D_0 \pi^b +\bar{c}_{ab} (D_0 \pi^a)^* D_0 \pi^b  \la{L}  \\ 
   &&  + d_{\alpha\beta} D_j \pi^\alpha D^j \pi^\beta  + d_{a \alpha} D_j \pi^a D^j \pi^\alpha +d_{ab} D_j \pi^a D^j \pi^b +\bar{d}_{ab} (D_j \pi^a )^* D^j \pi^b+ \mbox{c.c.}  \bigg\}, \nonumber 
\ea
where, for later convenience, we combined the fields $\pi_a^+$ and $\pi_a^-$ into a single complex field $\pi^a \equiv \pi^a_+ + i \pi^a_-$ with covariant derivatives
\be \la{cd4}
D_j \pi_a \simeq \partial_j \pi_a, \qquad \qquad D_0 \pi_a \simeq  \dot{\pi}_a - i \mu q_a \pi_a.
\ee

A few remarks are in order. First, the coefficients in the action are in general not all arbitrary, as they must be chosen in such a way as to make the Lagrangian  invariant under all {\it unbroken} symmetries. Second, as mentioned above, the action does not contain a term linear in $D_0 \pi_a$, because we require that there be no tadpoles. Finally, and perhaps more importantly, we should comment on the strong coupling scale of the low-energy effective theory.

\subsection{Strong coupling scale} \label{sec:scs}
When we expand the covariant derivatives in the action above to higher orders in the Goldstone fields, or when we add terms with higher powers of covariant derivatives, we will generate interaction terms. As befits a theory of Goldstone bosons, all such interactions are non-renormalizable, and, as a consequence, get strongly coupled in the UV, at some energy scale $\Lambda_{\rm strong}$. Usually this is not a problem, since one can work at energies that are far smaller than $\Lambda_{\rm strong}$, where the derivative expansion provides a  perturbative series that is well behaved at arbitrarily high orders. Our case, however, is subtler because, as we will confirm below, some of our Goldstones  have a gap of order $\mu$. If we want to include these modes consistently in the low-energy effective theory, we have to make sure that the strong coupling scale is well above $\mu$, 
\be \label{strong}
\Lambda_{\rm strong} \gg \mu \; .
\ee
We will now argue that this is a consistent assumption, but it is nonetheless an assumption, in the sense that there physical systems that do not obey it (while others do).

Consider first a relativistic theory that features SSB in the standard sense, that is, whose {\it Poincar\'e invariant} vacuum breaks some of the symmetries of the dynamics. There will be exactly massless Goldstone bosons, one for each broken generator, whose interactions get strongly coupled at some UV energy scale $f$.  This scale can be identified with the symmetry breaking scale. Consider now turning on a very weak density, or chemical potential, for one of the broken charges, with $\mu \ll f$. In terms of the Goldstone fields $\pi_a(x)$, this can be thought of as giving a time-dependent background to one of them, of the form $\pi_1(x) = \mu \, t$. Such a construction is carried out explicitly in \cite{Nicolis:2011pv}. This achieves our symmetry breaking pattern of sect.~\ref{setup}: on top of the symmetries that were already broken by the vacuum, the new state breaks Lorentz invariance, time-translations, and all the internal symmetries that do not commute with the charge associated with $\pi_1$.  The new Goldstone excitations will be described by our action \eqref{L}. 

Some of the Goldstones will now be gapped, with a gap of order $\mu$. However, since the strong coupling scale of the original Goldstone theory was $f$ and since, for $\mu \ll f$, the background Goldstone field we turned on {\it can} be described consistently within such an effective theory, we reach the unsurprising conclusion that the new effective action for the Goldstone excitations is nothing but the old one, expanded about the new background field. In particular, the strong coupling scale is still $f \gg \mu$ while  $\mu$ just plays the role of an infrared scale, which affects the low-energy spectrum, but has no important consequences at the level of interactions.  All interactions are still suppressed by inverse powers of $f$. Roughly speaking, the small scale $\mu$ always appears ``at the numerator," thus suppressing certain Lagrangian terms, rather than enhancing them. This  shows that eq.~\eqref{strong} is a consistent assumption.

Now consider instead a system in which there is no SSB in the absence of a chemical potential, for instance, liquid helium.  Helium atom number is spontaneously broken, i.e., helium atoms undergo Bose-Einstein condensation, only when there are helium atoms around, i.e., at finite density. For such a system the role of the symmetry breaking scale is played by the chemical potential, which also controls  the strength of the Goldstone interactions. To see this explicitly, consider for simplicity a relativistic superfluid, with a phonon speed of order of the speed of light, so that we don't need to differentiate between time- and space-derivatives. To lowest order in the derivative expansion but to all orders in the Goldstone field $\pi$, the low-energy effective action is~\cite{son}
\be \label{P(X)}
{\cal L} = P\big( \sqrt{(\dot \pi - \mu)^2 -(\partial_i \pi)^2} \, \big) \; .
\ee
Here $P$ is the same function that gives, at equilibrium (i.e., for vanishing $\pi$), the pressure as a function of the chemical potential. If one now expands this Lagrangian in powers of $\pi$,  assuming no hierarchy among the various derivatives of $P$, 
\be
P^{(n)}(\mu) \sim \mu^{4-n} \; , \qquad n=0,1,2, \dots
\ee
it is  clear that all Goldstone interactions will be weighed by the only scale characterizing the system, $\mu$, which is then the strong-coupling scale of the low-energy effective theory.

We thus reach the conclusion that both cases, i.e., $\Lambda_{\rm strong} \gg \mu$ and $\Lambda_{\rm strong} \sim \mu$, are consistent and physically relevant. As to our action \eqref{L}, the case with $f \sim \mu$ is clearly characterized by only one scale, which thus serves the role of the strong coupling scale as well,
\be
f \sim \mu \quad \Rightarrow \quad \Lambda_{\rm strong} \sim \mu \; .
\ee
The case with $f \gg \mu$ is more weakly coupled, but in general the strong coupling scale is not as high as $f$. This is because of the inverse powers of $\mu$ that are carried (via eq.~\eqref{IHeta}) by $\eta^i$, for instance in $\Lambda_\nu {}^\mu (\eta)$. It is easy to convince oneself that the strong coupling scale in this case is a geometric average of $f$ and $\mu$:
\be
f \gg \mu \quad \Rightarrow \quad \Lambda_{\rm strong} \sim \sqrt{\mu f} \; .
\ee
The simplest scenario where this happens is provided by the Lagrangian above for a relativistic superfluid, eq.~\eqref{P(X)}, with a function $P$ characterized by two scales:
\be \label{Ptilde}
P(\mu) = f^4 \tilde P(\mu/f) \; , 
\ee
where $\tilde P$ is a dimensionless function with order unity Taylor coefficients.

To actually end up with a case similar to the first example we analyzed above---where the turning on of  a small chemical potential in a very weakly-coupled Goldstone effective action had no effect on the strong-coupling scale of the theory---we need to choose a very special structure in our effective Lagrangian \eqref{L}. Without going into details for the general case, if we specialize to eq.~\eqref{P(X)} we see that by choosing
\be \label{very weak}
P(\mu) = f^4 \tilde {\tilde{ P}}\big( \mu^2/ f^2 \big) \; ,
\ee
where $\tilde {\tilde{ P}}$ is regular for its argument going to zero and has generically Taylor coefficients of order one, one gets rid of the square root structure in \eqref{P(X)}. As a result, when expanding in powers of $\pi$, one does not get any inverse powers of $\mu$. The $\mu \to 0$ limit is then regular by assumption, and one can think of the $\mu \neq 0$ case as a weak, infrared deformation of that. One thus gets that the strong coupling scale of this theory is $f$:
\be
f \gg \mu, \quad \mbox{no square root} \quad \Rightarrow \quad \Lambda_{\rm strong} \sim f \; .
\ee

To select this most weakly coupled structure directly at the level of the general low-energy effective theory \eqref{L} is trickier, but one thing is certain: such a structure is  technically natural, i.e., it is not destabilized by quantum corrections. The fundamental reason is that it corresponds to making interactions as weak as possible, and quantum loop corrections to all Lagrangian parameters are going to be suppressed accordingly. In detail, such a choice corresponds to having $f$ in the denominator in interaction terms, and $\mu$ appearing only in numerators, thus playing the role of coupling constants and mass parameters. The renormalization of Lagrangian coefficients involves the UV divergences of loop integrals, which are {\it analytic} in the tree-level coupling constants and mass parameters. In other words, if at the tree-level $\mu$ only appears at the numerator, loop corrections are going to keep it there.

We can get a sense of the parameter choices involved to select this most weakly coupled version of \eqref{L} by considering the $\mu \to 0$ limit. By demanding that the theory be weakly coupled at energies that are parameterically higher than $\mu$, we are effectively demanding that, at fixed energy, the theory have a smooth $\mu \to 0 $ limit. In particular, the number of Goldstones should be conserved in that limit, and they should all become gapless  {\it Lorentz scalars.} That is, under our assumption that Lorentz-boosts are only broken by the chemical potential, in the $\mu \to 0$ limit we should restore boosts and thus end up with an ordinary Lorentz-invariant theory of scalar fields. This constrains the $\mu \to 0$ limit of the Lagrangian coefficients.

By direct substitution of~\eqref{covdev} into~\eqref{L}, we find that in the $\pi$ sector Lorentz-invariance is recovered at $\mu = 0$ if and only if 
\begin{equation}
c(\mu = 0) = \sfrac{1}{2} \, b(\mu = 0) \, .
\end{equation}
As a check, notice that the simple template \eqref{very weak} obeys this constraint.
Analogous arguments for the whole $\pi$-$\pi_\alpha$  sector lead to 
\be
c_\alpha(\mu = 0)  = b_\alpha(\mu = 0) \; , \qquad d_{\alpha \beta} (\mu = 0) = c_{\alpha \beta} (\mu = 0) \; .
\ee

The broken non-commuting sector $\pi^a_\pm$ (or complex $\pi^a$) deserves more care. At first glance, the relation between the $c$'s and the $d$'s generalize to all sectors, simply giving $d_{\textsf{a} \textsf{b}} = c_{\textsf{a} \textsf{b}}$ at $\mu = 0$. However, as discussed at length in Sec.~\ref{sec:pipm}, depending on the unbroken symmetries, for some of the $(\pi^a_+, \pi^a_-)$ pairs we can choose gauges that get rid of one of the two fields---let's choose, conventionally, to keep $\pi^a_-$ and possibly get rid of $\pi^a_+$. This choice has to be made {\it before} taking  $\mu $ to zero, with implications for the coefficients of action~\eqref{L} that vary from case to case and can be quite intricate. For example, if we decide to get rid of $\pi^a_+$, by eq.~\eqref{after IH}, the time derivatives of $\pi^a_-$ will appear in those terms of the action containing $D_0\pi^a_+$, while the spatial kinetic terms still appear in $D_j \pi^a_-$. Thus, if we want to ensure that $\pi^a_-$ has a relativistic dispersion relation in the $\mu \rightarrow 0 $ limit, the pairing between time and space derivatives should be done accordingly.  

For finite but small $\mu$, we expect all these constraints on the Lagrangian parameters to be {\it almost} obeyed, with corrections suppressed by powers of $\mu/f$.

\section{Spectrum of Goldstones} \la{sec:spectrum}
In this section we derive the spectrum of our low-energy effective action (\ref{L}).  We find the four classes of Goldstone bosons described in the Introduction and we derive a counting rule for each of these types.  We compare our results to those in the previous literature.

\subsection{Scaling arguments}
The effective action (\ref{L}) is admittedly quite complicated and, at first sight, extracting any information about the spectrum of Goldstone bosons would seem hopeless. In fact, the standard procedure to derive the dispersion relations would require us to rewrite the quadratic part of the action in Fourier space,
\be
S_2 = \int \fr{d \omega \, d^3 k}{(2 \pi)^4} \l( \begin{array}{c} \pi \\ \pi^\textsf{a} \end{array} \r)^\dag \mbf{D} (\omega, k) \l( \begin{array}{c} \pi \\ \pi^\textsf{a} \end{array} \r),
\ee
and to solve the equation 
\be \label{secular}
\det \mbf{D} (\omega, k)= 0
\ee
for $\omega$. Fortunately, in most physical applications one does not need to know the exact form of the dispersion relations. This is because the infrared  behavior of the system is 
determined by the form  the dispersion relations take in the $k \to 0$ limit. Since $\mbf{D} (\omega, k)$ contains only terms of the form $\mu^2, \mu \, \omega, \omega^2$, and $k^2$, in the absence of fine-tunings we expect $\det \mbf{D}$ to vanish only for values of $\omega$ such that  $\mu^2 \sim \mu \, \omega \sim \omega^2$, or $ \omega^2 \sim k^2$, or $ \mu \, \omega \sim k^2$. In other words, in the absence of fine-tunings the spectrum will only contain gapped modes with $\omega \sim \mu$, linear gapless modes with $\omega \sim k$, and quadratic gapless modes with $\omega \sim k^2 /\mu$. In this section, we are going to derive counting rules for these three kinds of modes. 

To this end, we will use the fact that the total number of gapless modes can be deduced from the behavior of $\mbf{D} (\omega, 0 )$ in the   $\omega \to 0$ limit. In detail, we have
\be \la{lim1}
\lim_{\omega \to 0} \det [\mbf{D} (\omega, 0 ) ] \sim \omega^{2(n_1 + n_2)},
\ee
where $n_1$ and $n_2$ are respectively the number of linear gapless and quadratic gapless modes. This result can be easily checked when $ \mbf{D} (\omega, k)$ is diagonal, but remains valid in any basis, essentially by definition: the number of gapless modes {\it is} the number of $\omega^2=0$ solutions featured by eq.~\eqref{secular}  when $k$ goes to zero.

In order to determine separately how many gapless modes have a linear or quadratic dispersion relation, we can approach the  infrared from a complementary direction, and consider the behavior  of $\mbf{D} (0,k)$ in the  $k \to 0$ limit. In such a limit, linear and quadratic modes contribute differently to the scaling of $ \mbf{D} (0, k)$, and we get
\be \la{lim2}
\lim_{k \to 0} \det  [\mbf{D} (0, k) ] \sim k^{2(n_1 + 2 n_2)}.
\ee
Once again, this result can be checked in a basis where $ \mbf{D} (\omega, k)$ is diagonal, but is valid in any basis, for reasons analogous to the one above. By combining equations (\ref{lim1}) and (\ref{lim2}), one can easily solve for $n_1$ and~$n_2$. 

In section \ref{sec:pipialpha}, we will express $n_1$ and $n_2$ in terms of the structure constants of the internal symmetry group, thus showing that the number of linear and quadratic gapless modes is completely determined by the pattern of symmetry breaking. On the other hand, the number of gapped modes generically depends on how many gauge fixing conditions one chooses to impose.

 \subsection{Derivative mixings} \la{sec:mixing}
 
Our task of deriving the dispersion relations for the Goldstone modes is greatly simplified by the fact that the $\pi, \pi^\alpha$, and $\pi_\pm^a$ sectors  mix only via  operators that involve derivatives. In this section, we will derive some general results about derivative mixing that later on will allow us to study the $\pi_\pm^a$ sector separately.

Let us consider two sectors described by $\mbf{D}_1 (\omega, k)$ and $\mbf{D}_2 (\omega, k)$ respectively, and let us assume  their mixing to be  described by a mixing matrix $\mbf{M} (\omega, k)$. Based on the discussion in the previous section, we know that the numbers of linear and quadratic gapless modes are determined by the scaling properties of
\be\la{detDmix}
\det ( \mbf{D} ) \equiv \det \l( \begin{array}{cc} \mbf{D}_1 & \mbf{M}^\dag \\ \mbf{M} & \mbf{D}_2 \end{array} \r) = \det ( \mbf{D}_1 -  \mbf{M}^\dag  \mbf{D}_2^{-1}  \mbf{M} ) \det ( \mbf{D}_2 ) \, ,
\ee
in the infrared. The ``deconstruction" of the determinant follows from standard linear algebra results (see, e.g., \cite{det}). Now, the question we are interested in is the following: under what assumptions on $\mbf{M}$ is the number of linear and quadratic gapless modes {\it unaffected} by the mixing?

Let us start by considering the case in which the modes in  sector 2  are all gapped. Then, the ``mass matrix" $ \mbf{m}_2 \equiv \mbf{D}_2 (0, 0)$ is non-degenerate, i.e., $\det \mbf{m}_2 \neq 0$, and  can thus be inverted. It  follows from equation (\ref{detDmix})~that
\begin{subequations} \la{mix1}
 \ba
 \lim_{\omega \to 0} \det [\mbf{D} (\omega, 0 ) ] & \sim & \lim_{\omega \to 0} \det [ \mbf{D}_1  (\omega, 0 ) -  \mbf{M}^\dag  (\omega, 0 ) \cdot \mbf{m}_2^{-1}  \cdot \mbf{M}  (\omega, 0 ) ]  \\
 \lim_{k \to 0} \det [\mbf{D} (0, k) ] & \sim & \lim_{k \to 0} \det [ \mbf{D}_1  (0, k) -  \mbf{M}^\dag  (0, k) \cdot \mbf{m}_2^{-1} \cdot \mbf{M}  (0, k) ] \; .
 \ea
 \end{subequations}
Now, {\it if} $\mbf{M}  (\omega, k) \le \mathcal{O} (\omega,k)$, the  $\mbf{M}^\dag \mbf{m}_2^{-1} \mbf{M}$ terms on the RHS's of equations (\ref{mix1}), will at most change the coefficients in front of the terms proportional to $\omega^2$ and $k^2$, respectively in $\mbf{D}_1  (\omega, 0 )$ and $\mbf{D}_1  (0,k)$. However, in the absence of fine-tunings this will not change the overall scaling of the determinants in the  $\omega \to 0$, $k \to 0$ limits.
We therefore conclude that 
 \ba
 & \displaystyle \lim_{\omega \to 0} \det [\mbf{D} (\omega, 0 ) ] \sim  \lim_{\omega \to 0} \det [ \mbf{D}_1  (\omega, 0 ) ], &\\
 & \displaystyle \lim_{k \to 0} \det [\mbf{D} (0, k) ] \sim  \lim_{k \to 0} \det [ \mbf{D}_1  (0, k) ]. &
 \ea
This means that, barring fine-tunings, a one-derivative mixing between two sectors does not change the overall numbers of linear  gapless and quadratic gapless modes, as long as one of the two sectors  only features gapped excitations.

Let us now consider the case in which the sector 2 contains both gapped and gapless modes. Given  that the effective action (\ref{L}) contains at most two derivatives, each entry of the matrix $ \mbf{D}_2 (\omega, k)$ contains only terms proportional to $\mu^2, \mu \, \omega, \omega^2$ and~$k^2$. Furthermore, the determinant of $ \mbf{D}_2 (\omega, k)$ is given by 
\be
\det \mbf{D}_2 = \sum_{i} (-)^{i+j} \mbf{D}_2^{ij} \det \mbf{D}^{(ij)}_2
\ee
where we denoted with $\mbf{D}^{(ij)}_2$ the $(i,j)$ minor of $\mbf{D}_2$. Barring fine-tunings or accidental cancellations, we expect each term in the sum on the RHS to scale at least as fast as $\det \mbf{D}_2$ in the  $\omega \to 0$ and $k \to 0$ limits (eqs.~\eqref{lim1} and \eqref{lim2}). Since the elements of the inverse matrix $\mbf{D}_2^{-1}$ are 
\be
(\mbf{D}^{-1}_2)^{ij} = (-)^{i+j} \fr{\det \mbf{D}^{(ji)}_2}{\det \mbf{D}_2},
\ee
it follows that the nonzero entries of $\mbf{D}_2^{-1} (\omega, 0)$ and $\mbf{D}_2^{-1} (0,k)$ diverge at most as $1/ \omega^2$ and $1/ k^2$. {\it If} $\mbf{M} (\omega, k) \le \mathcal{O} (\omega^2, k^2)$, the term $\mbf{M}^\dag \mbf{D}_2^{-1} \mbf{M}$ will once again only change the coefficients of the  $\omega^2$ and $k^2$ terms in $\mbf{D}_1  (\omega, 0 )$ and $\mbf{D}_1  (0,k)$, but not the overall scaling of the determinant. Therefore, we have
 \ba
 & \displaystyle \lim_{\omega \to 0} \det [\mbf{D} (\omega, 0 ) ] \sim  \lim_{\omega \to 0} \det [ \mbf{D}_1  (\omega, 0 ) ] \det [ \mbf{D}_2  (\omega, 0 ) ], &\\
 & \displaystyle \lim_{k \to 0} \det [\mbf{D} (0, k) ] \sim  \lim_{k \to 0} \det [ \mbf{D}_1  (0, k) ] \det [ \mbf{D}_2  (0,k) ]. &
 \ea
and we conclude that a two-derivative mixing between two sectors  both containing gapless and gapped modes will not change the overall numbers of linear and quadratic gapless modes.
 
\subsection{$\pi^a$ sector} \la{sec:pipm}

Let us start by considering the action for the complex  $\pi^a$ field, without including its mixings  with the  $\pi$ and $\pi^\alpha$ fields. From equations (\ref{covdev}) and  (\ref{L}), we get:
\ba \la{piaaction}
S &=& f^2 \int d^4 x \Big\{ c_{ab} D_0 \pi^a D_0 \pi^b +\bar{c}_{ab} (D_0 \pi^a)^* D_0 \pi^b +d_{ab} D_j \pi^a D^j \pi^b  \nonumber  \\
&& \qquad\qquad\qquad   +\bar{d}_{ab} (D_j \pi^a )^* D^j \pi^b  + \mu_X \bar{b}_{ab} \pi^b (D_0 \pi^a)^* +  \mbox{c.c.}  \Big\}, 
\ea
where we have defined
\be
\bar{b}_{ab} \equiv - \fr{1}{4} \l[ (b f_{Xa_+ b_+} + b^\alpha f_{\alpha a_+ b_+}) - i (b f_{Xa_+ b_-} + b^\alpha f_{\alpha a_+ b_-}) \r].
\ee
After switching to Fourier space, we find that the inverse propagator for the doublets $(\pi_a, \pi_a^*)$ is equal to the following block matrix:
\ba
&\mbf{D} (\omega, k) \equiv \mbf{D}_{ \omega^2}  \, \omega^2+ \mbf{D}_{\omega} \, \mu  \omega + \mbf{D}_0 \, \mu^2 + \mbf{D}_{k^2} \,  k^2 \stackrel{k \to 0}{\longrightarrow}& \\
\nonumber \\
 &\l( \begin{array}{cc}
c_{ab} (\omega - \mu q_a) (\omega + \mu q_b)  & \bar{c}_{ab}^* (\omega - \mu q_a) (\omega - \mu q_b) + i \mu_X \bar{b}_{ab}^* (\omega - \mu q_a)\\  \bar{c}_{ab} (\omega + \mu q_a) (\omega + \mu q_b) + i \mu_X \bar{b}_{ab} (\omega + \mu q_b)  & c_{ab}^* (\omega + \mu q_a) (\omega - \mu q_b)
\end{array} \r)&\nonumber
\ea
where {\it no} sum over $a$ and $b$ is understood: for each pair of values for $a$ and $b$, there is a two-by-two block in $\mbf{D}$ that, in the limit $k \to 0$, takes precisely the form above.
The kinetic matrix $\mbf{D}_{\omega^2}$ must be non-degenerate, i.e. $\det (\mbf{D}_{\omega^2}) \neq 0$, in order for all the modes associated with $\pi_a$ to describe positive energy excitations. Then, it is easy to show that the mass matrix is also non-degenerate, because if we set $\bar{b}_{ab} =0$ we get
\ba
 \l.\det (\mbf{D}_0)\r|_{\bar{b}_{ab} =0} &=& \det  \l( \begin{array}{cc}
q_a \delta_{ac}  & 0 \\ 0 & - q_a \delta_{ac}
\end{array} \r) \l( \begin{array}{cc}
c_{cd} & \bar{c}_{cd}^* \\ \bar{c}_{cd} & c_{cd}^* 
\end{array} \r)  \l( \begin{array}{cc}
- q_b \delta_{db}  & 0 \\ 0 & q_b \delta_{db}
\end{array} \r) \\ 
&=&  \l[ \prod_a q_a^4 \r] \det \l( \begin{array}{cc}
c_{ab} & \bar{c}_{ab}^* \\ \bar{c}_{ab} & c_{ab}^* 
\end{array} \r) =  \l[ \prod_a q_a^4 \r] \det (\mbf{D}_{\omega^2}) \neq 0.
\ea
If we now turn back on $\bar{b}_{ab}$, we still expect that, in the absence of fine-tunings, all the modes in this sector will have a gap of order $\mu$. 

But we can go further: some of the modes have a gap that does not depend on any of the arbitrary coefficients appearing in the action (\ref{piaaction}). This follows from the fact that whenever $\omega = \pm \mu q_a$ for some $a$, one row or one column of $\mbf{D} (\omega, 0)$ vanish, and therefore $\det [ \mbf{D} (\omega, 0) ]= 0$. This means that for each complex field $\pi_a$, there is always a mode with gap 
\be
\omega^2 \to (\mu q_a)^2 \; , \qquad \mbox{for } k\to 0 \; ,
\ee 
which is completely determined by symmetry, since the $q_a$'s are related to the structure constants by equation (\ref{block}). 

The existence of these fixed gap modes follows from $\dot{\pi}_a$'s entering the action (\ref{piaaction}) only via the  combination $D_0 \pi_a = (\partial_t - i \mu q_a) \pi_a$. This remains true even when we allow for mixings with $\pi$ and $\pi_\alpha$, like in the action (\ref{L}).  When we vary w.r.t.~$\pi^a$ to obtain the equations of motion, we get the operator $ -(\partial_t + i \mu q_a)$ acting on whatever was multiplying $D_0 \pi_a$ in the action, thus always allowing for an $\omega = \mu q_a$ solution. (The other solution, with $\omega = - \mu q_a$, comes from considering the $\pi_a^*$ field).
This proves that for each pair of broken generators that do not commute with the charge at finite density, there is always a mode with fixed gap $\omega= |\mu q_a|$, in agreement with what was found in~\cite{Nicolis:2012vf}. More precisely, the number $n_3$ of the modes with fixed gap is given by the rank of the matrix $M_{\textsf{a}\textsf{b}}$ defined in equation (\ref{block}) or, directly in terms of the structure constants, by
\be \la{n3}
n_3 = \tfrac{1}{2} \mbox{rank} (\mu_X f_{X \textsf{a}\textsf{b}} + \mu_T f_{T \textsf{a}\textsf{b}}).
\ee

Since any complex field such as $\pi^a$ describes two degrees of freedom, one may be tempted to conclude that for each mode with fixed gap $m_a = \mu q_a$ there is always a partner mode with gap $m \sim \mu$. The situation is however a bit subtler, since one can in principle reduce the number of degrees of freedom in this sector by imposing some gauge fixing conditions. This follows from the commutation relations between the unbroken Hamiltonian $\bar{P}_0$ and the generators associated with $\pi^a = \pi^a_+ + i \pi^a_-$:
\be
[ \bar{P}_0, X_a^\pm ] = \pm i  \mu q_a X_a^\mp.
\ee
{\it If} $\pi^a_+$ and $\pi^a_-$ do not transform in the same irreducible multiplet under the unbroken symmetries, one can choose to impose the constraint $D_0 \pi^-_a = 0$ and  express  $\pi^a_+$ in terms of $\pi^a_-$:\footnote{Equivalently, one could choose to remove $\pi^a_-$ by imposing $D_0 \pi^+_a = 0$. The two choices are formally different but physically equivalent, because they are simply two different gauge choices for the same gauge redundancy.}
\be
\pi_a^+ = \fr{\d_0 \pi_a^-}{\mu q_a} + \mathcal{O}( \pi_-^2) \; .
\ee
After choosing such a gauge, the quadratic part of the action (\ref{piaaction}) at zero momentum depends on $\pi_a^-$ only through the combinations
\be \label{after IH}
D_0 \pi^+_a =  \fr{\d_0^2 \pi_a^-}{\mu q_a} + \mu q_a \pi_a^- + \mathcal{O}( \pi_-^2) \; .
\ee
Thus, the gauge choice $D_0 \pi^-_a = 0$ does not affect the mode with fixed gap $m_a = \mu q_a$---varying eq.~\eqref{after IH} w.r.t.~$\pi^-_a$ yields the differential operator $(\partial_t^2+(\mu q_a)^2)$, 
which still allows for the $\omega= \pm \mu q_a$ solutions---but removes its gapped partner from the spectrum. Based on our discussion in section \ref{IH}, we conclude that the existence of the first mode follows solely from the pattern of symmetry breaking, whereas the existence of its gapped partner is more model-dependent, or, more precisely, system-dependent. In section \ref{sec:so(3)} we are going to illustrate this point with explicit examples.

\subsection{$\pi, \pi^\alpha$ sector} \la{sec:pipialpha}

The modes $\pi$ and $\pi^\alpha$ have a one-derivative mixing with the $\pi^a$ sector, which as we have seen only contains gapped modes. Based on the general results of section \ref{sec:mixing}, for the purpose of determining the number of linear gapless and quadratic gapless modes, we can therefore neglect all such mixings and study the modes $\pi,\pi^\alpha$ separately. Their quadratic action takes the form:
\ba \la{S3}
S &=& f^2 \int d^4 x \l[ c \, \dot{\pi}^2 -\frac{b}{2} \, \d_j \pi \d^j \pi + \mu_X b^\alpha f_{X\alpha\beta} \dot{\pi} \pi^\beta - b_\alpha \d_j \pi \d^j \pi^\alpha + c_{\alpha\beta} \dot{\pi}^\alpha \dot{\pi}^\beta \r. \\
&& \qquad\qquad\qquad\qquad\qquad\qquad \l.+ \frac{\mu_X}{2}  ( b^\alpha  f_{\alpha\beta\gamma} + b f_{X\beta\gamma}) {\pi^\beta \dot{\pi}^\gamma} + d_{\alpha\beta} \d_j \pi^\alpha \d^j \pi^\beta \r]. \nonumber 
\ea
Based once again on the discussion in section \ref{sec:mixing}, we are going to neglect all the two-derivative mixings in the action (\ref{S3}). Then, the number of linear, quadratic and massive Goldstones in this sector is crucially determined by the mixing matrix
\be
\mbf{M} = \l(\begin{array}{c | c} 0 & - b^\gamma f_{X\gamma\beta} \\ \hline \\ b^\gamma  f_{X\gamma\alpha} \,\, & \,\,  b^\gamma  f_{\gamma\alpha\beta} + b f_{X\alpha\beta}\\ &  \end{array} \r).
\ee
Since $\mbf{M}$ is an antisymmetric matrix, we can always cast it into a block diagonal form:
\be \la{block2}
\mbf{M} =  \mbox{diag} \l\{ 0, \cdots, 0, 
\l(
\begin{array}{cc}
0 &M_1\\
- M_1 & 0 
\end{array}
\r), \cdots, 
\l(
\begin{array}{cc}
0 &M_k\\
- M_k & 0 
\end{array}
\r)\r\}
\ee
Then, for each zero on the diagonal in the RHS of equation (\ref{block2}) we get one linear gapless Goldstone. This means that the total number of linear gapless modes is
\be \la{nlin}
n_1 = \dim (\mbf{M}) - \mbox{rank} \, (\mbf{M}),
\ee
On the other hand, each pair of fields corresponding to a two-by-two block in equation (\ref{block2}) has an inverse propagator of the  form
\be
\mbf{D} (\omega, k) = \l( \begin{array}{cc} 
c_1^2 \omega^2 - d_1^2 k^2 \quad & \sfrac i2 M \mu_X \, \omega  \\ - \sfrac i2 M \mu_X \, \omega  \quad & c_2^2 \omega^2 - d_2^2 k^2
\end{array} \r).
\ee
By setting $\det [\mbf{D} (\omega, k) ] =0$ and solving for $\omega$, we get a gapped mode with $k \to 0$ dispersion relation $\omega \sim \mu$, and a quadratic gapless one with dispersion relation $\omega \sim k^2 / \mu$. Since this sector is the only one containing  quadratic gapless modes, their total number is simply
\be\la{nquad}
n_2 = \sfrac{1}{2} \, \mbox{rank} \, (\mbf{M}).
\ee

Notice that for each quadratic gapless mode, there is always an associated unfixed-gap mode with $\omega \sim \mu$. Such a mode is never redundant, that is, it can never be eliminated by a gauge fixing condition, because in this sector we have trivial commutators with the unbroken Hamiltonian:
\be
[\bar P^0, X_\alpha] = 0 \; .
\ee 
We thus see that the total number of unfixed-gap Goldstones, $n_4$, cannot be smaller than the number of gapless-quadratic ones, $n_2$. In addition to this, there are in general the unfixed-gap partners of the fixed-gap Goldstones discussed in sect.~\ref{sec:pipm}, whose existence and actual number is more model-dependent, and of which we have at most as many as the number of fixed-gap Goldstones, $n_3$. We thus have the bound
\be \label{n4}
n_2 \le n_4 \le n_2 + n_3 \; .
\ee

Equations \eqref{n3}, (\ref{nlin}), (\ref{nquad}), and \eqref{n4} are among the main results of this paper, because they allow us to derive the number of Goldstone modes of each type from the algebra of the internal symmetry group and the symmetry breaking pattern.

\subsection{Comparison with previous literature}
Although our results about the existence and number of unfixed-gap Goldstones are entirely new, our counting rules for $n_1$, $n_2$, and $n_3$ almost completely agree with those recently proposed in~\cite{Watanabe:2013uya} (see also~\cite{Schafer:2001bq,Watanabe:2011ec,Watanabe:2012hr,Hidaka:2012ym}). 
There is however  an apparent small discrepancy, whose physical relevance we are going to comment about below.

The counting rule for quadratic gapless modes derived in~\cite{Watanabe:2013uya} is, in our notation,
\begin{subequations}\la{count1}
\ba
n'_2 &=& \lim_{V \to \infty} \fr{1}{2 V} \, \mbox{rank} \l(\begin{array}{c | c} 0 & \langle\mu | [X,X_\beta] | \mu \rangle \\ \hline \\ \langle\mu | [X_\alpha,X] | \mu \rangle \,\, & \,\,  \langle\mu | [X_\alpha,X_\beta] | \mu \rangle \\ &  \end{array} \r) \\
&=& \sfrac{1}{2} \mbox{rank}  \l(\begin{array}{c | c} 0 & -  \langle j^\gamma_0 \rangle f_{X\gamma\beta} \\ \hline \\  \langle j^\gamma_0 \rangle  f_{X\gamma\alpha} \,\, & \,\,  f_{\alpha\beta\gamma} \langle j^\gamma_0 \rangle + f_{\alpha\beta X} \langle j^X_0 \rangle + f_{\alpha\beta T} \langle j^T_0 \rangle\\ &  \end{array} \r). \la{count1b}
\ea
\end{subequations}
where $V$ is the spatial volume, and in the last step we used that $f_{\alpha\beta a_\pm} $ vanishes, and assumed that $T$ is the only unbroken charge at finite density (given that it is the only one for which there is a non-zero chemical potential). 

On the other hand, from our Goldstone action \eqref{L} we can immediately derive the tree-level contributions to the current densities for the {\it broken} charges $X_\alpha$ and $Q$:\footnote{These identities follow from the fact that, if the symmetries generated by the $X^\alpha$ and $X$ were gauged, the associated N\"other currents would be $j^\mu_\alpha =\delta S/ \delta A_\mu^\alpha$ and $j^\mu_X =\delta S/ \delta A_\mu^X$ where $A_\mu^\alpha$ and $A_\mu^X$ are the respective gauge fields. In our context these symmetries are only global, but the Goldstones $\pi$ and $\pi^\alpha$ can be thought of as St\"{u}ckelberg fields contributing to fictitious gauge fields as pure gauge components, $A^X_\mu = D_\mu \pi$, $A_\mu^\alpha = D \pi^\alpha$.}
\be \label{currents}
j^0_\alpha = \fr{\delta S}{\delta D_0 \pi^\alpha} = f^2 \mu_X \, b_\alpha + \mathcal{O}(\pi), \qquad\quad j^0_X = \fr{\delta S}{\delta D_0 \pi} = f^2 \mu_X \, b + \mathcal{O}(\pi) \; ,
\ee
so that our counting rule  (\ref{nquad}) can be rewritten as in equation (\ref{count1b}) but without the term proportional to the expectation value of the unbroken current $ \langle j^T_0 \rangle$.

When $\mu_T=0$, our counting rule agrees exactly with equation (\ref{count1b}) because $ \langle j^T_0 \rangle$ must vanish as well. Moreover, since $X$  coincides with $Q$ we have by definition $f_{X\alpha\beta} =0$ and thus the counting rule for $n_2$ becomes even simpler. When $\mu_T \neq 0$, we can use the symmetry algebra to relate the last two terms in \eqref{count1}: 
\be
\mu[Q, X_\alpha] = 0 \quad \Rightarrow \quad  f_{\alpha\beta T} = - \fr{\mu_X}{\mu_T} f_{\alpha\beta X} \; ,
\ee
Then, the lower right block in equations (\ref{nquad})  and (\ref{count1b})  clearly involve the same matrices, namely $f_{\alpha\beta\gamma} \langle j^\gamma_0 \rangle$ and $f_{\alpha\beta X}$, but not quite in the same combination. We still expect these two equations to give the same number of quadratic gapless modes for {\it generic} values of $\mu_X$ and $\mu_T$. We were however unable to prove that the two counting rules are equivalent for {\it all} values of the chemical potentials.

\section{Example: $SU(2) \times U(1) \to U(1)$}
\label{sec:su(2)} 
Let us now turn to some specific examples.  In this section we consider the breaking of $SU(2) \times U(1)$ down to $U(1)$.  In the case that the charge of the original $U(1)$ symmetry is at finite chemical potential, this system can be used as a simple model for Kaon condensation \cite{Schafer:2001bq,Miransky:2001tw}.  We first use the coset construction to obtain the generic spectrum of the theory.  We find one massless mode, one massive mode whose mass is fixed and one unfixed massive mode.  We then consider a specific UV realization of this symmetry breaking pattern in the form of a complex doublet.  We compare our results to that of the coset construction.

\subsection{The coset construction}
We wish to apply the formalism developed above to the case of $SU(2) \times U(1) \rightarrow U(1)$.  The symmetry generators are denoted by $L_i$ for the $SU(2)$ symmetry and $Y$ for the initial $U(1)$.  We consider a scenario in which the unbroken $U(1)$ is given by the combination $Y+L_3$.  Thus to guarantee that the structure constants are totally antisymmetric and the maximum number of unbroken generators appear, we choose the following basis:
\ba
\begin{array}{lcl}
\mbox{unbroken} &=&  \l\{
\begin{array}{ll}
\bar{P}_0 \equiv P_0 + \mu Q & \qquad \qquad \qquad \mbox{time translations} \\
\bar{P}_i \equiv P_i &  \qquad \qquad \qquad \mbox{spatial translations}  \\
J_i &  \qquad \qquad \qquad \mbox{rotations}\\
T \equiv \fr{1}{\sqrt{2}} (Y+L_3)  &  \qquad \qquad \qquad \mbox{internal symmetry}
\end{array}
\r. \nonumber 
\\
\\
\mbox{broken} &=&  \l\{
\begin{array}{ll}
K_i &  \qquad  \quad \!\!  \mbox{boosts} \\
L_1,\,L_2, \, X\equiv \fr{1}{\sqrt{2}} (Y-L_3) &  \quad  \qquad \!\! \mbox{internal symmetries}
\end{array}
\r. \nonumber 
\end{array}
\ea
where
\be
\mu Q =  \mu_X X + \mu_T T = \fr{\mu_X+\mu_T}{\sqrt{2}} \, Y-\fr{\mu_X-\mu_T}{\sqrt{2}} \,  L_3 \, .
\ee
We parametrize the coset as follows,
\be
\la{cosparSU}
\Omega = e^{i x^\mu \p_\mu} e^{i \pi X} e^{i \pi^a L_a} e^{i \eta^i K_i} \, ,
\ee
where $a=1,2$.  In the language that we have been using throughout, when $\mu_T\neq \mu_X$ the $L_a$'s make up a pair of non-commuting broken generators $X_\pm$, since $\mu [Q, L_{1,2}] = \pm i L_{2,1} (\mu_T - \mu_X) /\sqrt{2}$. Following the procedure given above, the covariant derivatives at lowest order are given by
\be
\label{DsSU}
\begin{array}{lcl}
D_0 \pi &\simeq& \dot{\pi} - \frac{1}{2\mu_X}\partial_j \pi \partial^j \pi+ \frac{1}{2\sqrt{2}}(\dot{\pi}_1 \pi_2-\pi_1\dot{\pi}_2)  + \frac{1}{4}(\mu_T-\mu_X)(\pi_1^2+\pi_2^2) \, ,\\ 
D_0 \pi_1 &\simeq&  \dot{\pi}_1+ \fr{\mu_T-\mu_X}{\sqrt{2}} \pi_2 \, ,\\
D_0 \pi_2 &\simeq&  \dot{\pi}_2 - \fr{\mu_T-\mu_X}{\sqrt{2}} \pi_1 \, ,\\
D_j \pi_1 &\simeq&  \partial_j \pi_1 \, , \\
D_j \pi_2 &\simeq&  \partial_j \pi_2 \, .
\end{array}
\ee
When $\mu_T = \mu_X$, the generators $L_a$'s become commuting generators $X_\alpha$, but the covariant derivatives above are correct even in this limit.

To construct the most generic Lagrangian that is manifestly invariant under the unbroken $Y+L_3$, we note that the $\pi_a$'s transform as a doublet under this symmetry and thus all $a$ indices should be contracted.  The quadratic Lagrangian can thus be expressed as,
\be
{\cal L}_2 = c_1 \, \mu_X D_0 \pi+c_2\,(D_0 \pi)^2+c_3\,(D_0\pi_a)(D_0\pi^a)+c_4\, (D_j \pi_a)(D_j\pi^a) \, .
\ee
Substituting in the covariant derivatives (\ref{DsSU}), it is straightforward to see that the $\pi$ modes decouple from the $\pi_a$'s.  The $\pi$ dispersion relation is given by
\be
\omega^2=\frac{c_1}{2c_2}k^2 \, .
\ee
This mode is clearly massless, regardless of the coefficients $c_i$.  The $\pi_a$'s mix with each other.  In the zero momentum limit $k\rightarrow 0$, we find the masses of these two modes to be,
\be
\label{omSU}
\begin{array}{lcl}
\omega_+ (k\rightarrow 0)&=&\l| \left(\frac{c_1}{2c_3}-1\right) \fr{\mu_X}{\sqrt{2}} +\fr{\mu_T}{\sqrt{2}} \r| \, \\
\omega_-(k\rightarrow 0) &=& \fr{1}{\sqrt{2}} |\mu_X-\mu_T|  \, .
\end{array}
\ee
The second mode has a mass that is independent of the coefficients $c_i$.  In the case that $\mu_X = \mu_T$, i.e, when $Q \sim Y$, this mode is exactly massless.  The first mode is generically massive, with a mass of order $\mu_X,\mu_T$.  While we can tune the $c_i$ coefficients to make this mode massless, there is no symmetry that protects this tuning.

Note that this theory has in principle a potential gauge-redundancy, since
\be
[\p_0,L_1] \sim L_2\, , ~~~~ [\p_0,L_2] \sim L_1\, .
\ee
These commutators would indicate that we could set either $D_0\pi_1=0$ or $D_0\pi_2=0$ as a gauge-fixing condition.  However, as the $\pi_a$'s transform as a doublet under the unbroken $Y+L_3$, we cannot impose either without violating this symmetry.  Thus the potential redundancy and its removal via fixing a gauge are inconsistent with the unbroken symmetries,  and all three Goldstones will appear as physical degrees of freedom in a theory with this symmetry breaking pattern.

In the following subsection we consider an explicit UV theory that realizes this symmetry breaking pattern and verify the spectrum of this theory against that of the coset.

\subsection{Linear sigma model for a complex doublet}
Let us consider the above symmetry breaking pattern realized by a complex doublet $\Phi$ of $SU(2)$,
\be
\label{LSU}
{\cal L} = (\partial_\mu \Phi)^\dag (\partial_\mu \Phi)-m^2 \Phi^\dag \Phi-\lambda(\Phi^\dag\Phi)^2 \, .
\ee
We introduce chemical potentials for the charges $Y-L_3$ and $Y+L_3$ via the standard replacement,
\be
\partial_0 \rightarrow \partial_0+i \mu_X \fr{(Y-L_3)}{\sqrt{2}}+i\mu_T \fr{(Y+L_3)}{\sqrt{2}} \, .
\ee
Let us also introduce the field redefinition,
\be
\Phi(x) = \frac{1}{\sqrt{2}} \left(\begin{array}{c}
\tilde{\phi}_1(x)+ i\tilde{\phi}_2(x) \\
v+\phi_1(x)+ i\phi_2(x)
\end{array}
\right)
\ee
The expectation value $v$ breaks the $Y-L_3$ symmetry and is given by
\be
v=\sqrt{\frac{\mu_X^2-2 m^2}{2 \lambda}} \, .
\ee
Around this background, the dispersion relations for the four fields $\phi_{1,2}$, $\tilde{\phi}_{1,2}$ are given by
\be
\begin{array}{lcl}
\omega_\pm^2 & = & k^2+ 3 \mu_X^2/2-m^2 \pm \sqrt{2k^2\mu_X^2+(3\mu_X^2/2-m^2)^2} \, , \\
\tilde{\omega}_\pm &=& \sqrt{k^2+\mu_X^2/2} \pm \mu_T/\sqrt{2} \, .
\end{array}
\ee
In the zero momentum limit these dispersion relations become
\be
\begin{array}{lcl}
\omega_+ (k\rightarrow 0)& = & \sqrt{3\mu_X^2-2m^2} \, , \\
\omega_- (k\rightarrow 0)& = & c_- \cdot k \, , \qquad c_- = \sqrt{\sfrac{\mu_X^2-2m^2}{3\mu_X^2-2m^2}}\\
\tilde{\omega}_+(k\rightarrow 0) &=& \fr{1}{\sqrt{2}} |\mu_X+ \mu_T| \, ,  \\
\tilde{\omega}_-(k\rightarrow 0) &=& \fr{1}{\sqrt{2}} |\mu_X- \mu_T| \, .
\end{array}
\ee
The dispersion relation $\omega_+$ is that of the radial mode and is not captured by the coset construction.  In a scenario where the symmetries of the theory are broken even in the absence of a chemical potential, one could take $m \rightarrow \infty$ while keeping $\mu$ finite and this mode with $\omega_+ \sim m $ would be beyond the scope of our low energy effective theory.  

The dispersion relation $\omega_-$ is for the massless Goldstone boson represented above in the coset construction by $\pi$, with the coefficients $c_i$ chosen so that $\frac{c_1}{2c_2} = c_-^2$.  The final two dispersion relations correspond to linear combinations of $\pi_1$ and $\pi_2$ of the coset construction.  The dispersion relation $\tilde{\omega}_+$ corresponds to the mode of unfixed mass given in (\ref{omSU}), with the coefficients $c_i$ chosen so that $\frac{c_1}{2c_3}-1 = 1$.  The dispersion relation $\tilde{\omega}_-$ corresponds precisely to the fixed mass Goldstone in (\ref{omSU}).  The coset construction tells us that this mass is determined entirely by the symmetry breaking pattern and thus we expect it not get corrected by quantum effects.\footnote{In fact, this particular choice of Lagrangian \eqref{LSU} has an extended $SO(4) \sim SU(2)_L \times SU(2)_R$ custodial symmetry, which can be used to exactly determine the gap of $\tilde{\omega}_+$ \cite{Watanabe:2013uya}.}

\section{Example: $SO(3) \to \emptyset$} \la{sec:so(3)}
In this section we consider a theory with an internal $SO(3)$ symmetry which is completely broken by the ground state, after the introduction of a chemical potential $\mu$ for one of the $SO(3)$ charges. In the language of Spontaneous Symmetry Probing (SSP), all generators are broken spontaneously by a time-dependent field configuration.  We consider this theory first using the coset construction.  We then consider three explicit UV theories that realize this symmetry breaking pattern.  We discuss the relevance of the Goldstone gauge redundancy to these theories.

\subsection{The coset construction}
We denote the internal $SO(3)$ generators by $L_i$. To describe $SO(3)$ breaking using the coset construction, we choose our basis of generators in the following way,
\ba
\label{patternSO}
\begin{array}{lcl}
\mbox{unbroken} &=&  \l\{
\begin{array}{ll}
\bar{P}_0 \equiv P_0 +\mu L_3 & \qquad \qquad \qquad \mbox{time translations} \\
\bar{P}_i \equiv P_i &  \qquad \qquad \qquad \mbox{spatial translations}  \\
J_i &  \qquad \qquad \qquad \mbox{rotations}
\end{array}
\r. 
\\
\\
\mbox{broken} &=&  \l\{
\begin{array}{ll}
K_i &  \qquad \qquad \qquad \qquad  \quad \!\!\!\!\!\!  \mbox{boosts} \\
L_1,\,L_2, \, L_3 &  \qquad  \qquad \qquad  \quad \qquad \!\!\!\!\!\! \mbox{internal symmetries}
\end{array}
\r.  
\end{array}
\ea
We parametrize the coset as follows,
\be
\Omega = e^{i x^\mu \p_\mu} e^{i \pi L_3} e^{i \pi^a L_a} e^{i \eta^i K_i} \, ,
\ee
where again $a=1,2$.  Now the covariant derivatives at lowest order are given by
\be
\label{DsSO}
\begin{array}{lcl}
D_0 \pi &\simeq& \dot{\pi} - \frac{1}{\mu}\partial_j \pi \partial^j \pi+\frac{1}{2}(\dot{\pi}_1 \pi_2-\pi_1\dot{\pi}_2)  -\frac{1}{2}\mu(\pi_1^2+\pi_2^2) \, ,\\ 
D_0 \pi_1 &\simeq&  \dot{\pi}_1+\mu\, \pi_2 \, ,\\
D_0 \pi_2 &\simeq&  \dot{\pi}_2 -\mu\, \pi_1 \, ,\\
D_j \pi_1 &\simeq&  \partial_j \pi_1 \, , \\
D_j \pi_2 &\simeq&  \partial_j \pi_2 \, .
\end{array}
\ee
As all internal symmetries are broken, the most general quadratic Lagrangian can be expressed as,
\be
\begin{array}{lcl}
{\cal L}_2 &=& c_1 \, \mu D_0 \pi+c_2\,(D_0 \pi)^2+c_3\,(D_0\pi)(D_0\pi_1)+c_4\,(D_0\pi)(D_0\pi_2)  \\
&&+c_5\,(D_0\pi_1)(D_0\pi_1)+c_6\,(D_0\pi_2)(D_0\pi_2) +c_7\,(D_0\pi_1)(D_0\pi_2)\\
&&+c_8\,(D_j\pi_1)(D_0\pi_1) + c_9\,(D_j \pi_2)(D_j\pi_2) + c_{10}\,(D_j \pi_1)(D_j\pi_2) \, .
\end{array}
\ee
All three modes mix with each other and the mixing is somewhat more involved than the case of $SU(2)\times U(1)$ breaking.  Nevertheless, in the zero momentum limit one finds three dispersion relations of the form,
\be
\begin{array}{lcl}
\omega(k\rightarrow 0) &=& 0 \, , \\
\omega_+(k\rightarrow 0) &=& \mu f(c_1,\ldots , c_7) \, , \\
\omega_-(k\rightarrow 0) &=& \mu \, .
\end{array}
\ee
The first mode is massless.  The second mode is generically massive, with a mass of order $\mu$ that depends on  the coefficients via a specific function $f(c_1,\ldots , c_7)$.  The third mode has a mass $\mu$ that  is fixed.

Similar to the case of $SU(2)\times U(1)$ breaking, this theory has a potential gauge redundancy, as
\be
[\p_0,L_1] \sim L_2\, , ~~~~ [\p_0,L_2] \sim L_1\, .
\ee
As per the usual logic \cite{Ivanov:1975zq}, these commutators indicate that it is possible to set either $D_0\pi_1=0$ or $D_0\pi_2=0$ as consistent gauge choice.  Since there are no unbroken internal symmetries, imposing these relations does not violate the symmetry breaking pattern, unlike the case of $SU(2)\times U(1) \rightarrow U(1)$.  In what follows, we will see that whether or not these relations should be imposed is model-dependent, in the sense that it depends on which $SO(3)$ representation the order parameter belongs to. We will now consider three UV Lagrangians that realize the pattern of symmetry breaking (\ref{patternSO}).

\subsection{Linear sigma model for one triplet}

The first example we are going to consider was discussed thoroughly in~\cite{Nicolis:2012vf}. Therefore, we will content ourselves with reviewing it very briefly and we will refer the reader to~\cite{Nicolis:2012vf} for a more detailed analysis. The simplest model we can consider that realizes the symmetry breaking pattern (\ref{patternSO}) is one that contains a single $SO(3)$ triplet $\phi^n$ described by the Lagrangian
\be
\mathcal{L} = -\fr{1}{2} \d_\mu \phi_n \d^\mu \phi^n - \fr{\lambda}{4} \l( \phi_n \phi^n - v^2\r)^2.
\ee
In this model, the SSP field configuration
\be
\langle \phi \rangle = e^{i \mu t L_3} \l( \begin{array}{c} \phi_0 \\ 0 \\  0 \end{array} \r) \,,
\ee
with $\phi_0 = \sqrt{v^2 + \mu^2 /\lambda}$ and $(L_3)_{ij} \equiv - i\epsilon_{3ij}$, is responsible for breaking $SO(3)$ completely. In this case, besides a radial mode with mass $m = \sqrt{\lambda} \, \phi_0$, the spectrum contains only one massless Goldstone with linear dispersion relation and one massive Goldstone with mass $m =\mu$. In particular, there is no other mode with mass $m \sim \mu$. In the limit $\lambda \phi_0^2 \gg \mu^2$ one can integrate out the radial mode to get a low-energy effective action for the Goldstone bosons. 

This is the same spectrum one finds from the coset construction if one gauge-fixing condition is imposed (in addition to the conditions necessary to eliminate the Goldstones associated with the broken boosts). Following the logic of section \ref{IH}, we can see why we have a gauge redundancy in the Goldstone parameterization of the system.  To start, note that even though the internal $SO(3)$ is completely broken, there is a {\it time-dependent} linear combination of internal generators that acting on $\langle \phi \rangle$ gives zero
\footnote{This is not the same as saying that such a combination is unbroken, for there are other non-vanishing expectation values, like for instance $\langle \do \partial_\mu \phi \rangle$, which are {\it not} annihilated by it. In the language of Sec.~\ref{inter} below, the combination $\bar L_1$ is ``non-interpolating".},
\be
\label{IHSO}
\bar{L}_1=e^{i\mu t L_3}L_1e^{-i\mu t L_3} = \cos(\mu t)\, L_1-\sin(\mu t)\, L_2\, , ~~~~\bar{L}_1 \langle \phi \rangle =0\, .
\ee
If we now consider fluctuations of the fields $\phi_n$, they are given at lowest order in the Goldstone fields as
\be
\delta \phi_n \simeq i (\pi_1 L_1+\pi_2 L_2+\pi_3 L_3) \langle \phi \rangle \, .
\ee
Using the relation (\ref{IHSO}), we see that $\delta \phi_n$ is invariant (at lowest order in fields) under the gauge transformation
\be
\begin{array}{c}
\pi_1 \rightarrow \pi_1+\Lambda(x)\cos(\mu t) \, ,\\
\pi_2 \rightarrow \pi_2-\Lambda(x)\sin(\mu t) \, ,
\end{array}
\ee
where $\Lambda(x)$ is any generic function of space and time.  This redundancy implies that $\pi_1$ and $\pi_2$ do not describe physically independent fluctuations.

To eliminate this redundancy we must pick a gauge.  The coset construction offers two choices for this gauge condition that transform covariantly under the desired symmetries: either $D_0 \pi_1=0$ or $D_0 \pi_2=0$.  Choosing the first condition will allow one to eliminate $\pi_2$ in favor of $\pi_1$, and vice versa for the second condition.  Since these are simply gauge choices, the physical content of the Lagrangian is the same for either choice.  After the gauge fixing condition is imposed in the coset construction, one arrives at the correct spectrum for the above UV theory.

\subsection{Linear sigma model for two triplets}

In our second example, we will break $SO(3)$ with two triplets $\phi$, $\tilde \phi$. In this model, $SO(3)$ remains completely broken even in the limit $\mu \to 0$. The most generic Lagrangian for two triplets which is invariant under $\mathbb{Z}_2$ symmetries acting separately on the two triplets is:
\ba \la{La2}
\mathcal{L} = -\sfrac{1}{2} \big(\d_\mu \phi\big)^2 - \sfrac{\lambda}{4} \big(\phi^2 - v^2\big)^2  -\sfrac{1}{2} \big(\d_\mu \tilde{\phi}\big)^2  - \sfrac{\tilde{\lambda}}{4} \big( \tilde{\phi}^2  - \tilde{v}^2\big)^2 - \sfrac{g}{2} (\phi \cdot \tilde{\phi} )^2 - \sfrac{\kappa}{2} \phi^2 \tilde{\phi}^2 \quad
\ea
We will assume that all coupling constants are positive and that $v^2 > \tilde{v}^2$. At $\mu =0$, the term proportional to $g$ forces $\phi$ and $\tilde{\phi}$ to have vacuum expectation values that are orthogonal to each other. We will therefore consider the following SSP field configuration:
\be \la{SPP2}
\langle \phi \rangle = e^{i \mu t L_3} \l( \begin{array}{c} \phi_0 \\ 0 \\  0 \end{array} \r), ~~~~  \langle \tilde{\phi} \rangle = e^{i \mu t L_3} \l( \begin{array}{c} 0 \\\tilde{\phi}_0 \\  0 \end{array} \r), \nonumber
\ee
with
\be
\phi_0 = \sqrt{\fr{\tilde{\lambda} (\lambda v^2 + \mu^2) - \kappa (\tilde{\lambda}\tilde{v}^2 + \mu^2 )}{\lambda \tilde{\lambda} - \kappa^2}} \, , \qquad 
 \tilde{\phi}_0 = \sqrt{\fr{  \lambda (\tilde{\lambda}\tilde{v}^2 + \mu^2 ) - \kappa  (\lambda v^2 + \mu^2) }{\lambda \tilde{\lambda} - \kappa^2}} \, . 
\ee
This field configuration minimizes the effective potential provided $\kappa$ is small enough, i.e. such that
\be
\kappa < \mbox{min} \l\{ \sqrt{\lambda \tilde{\lambda}}, \, \fr{\tilde{\lambda} (\lambda v^2 + \mu^2)}{ (\tilde{\lambda}\tilde{v}^2 + \mu^2 )}, \,  \fr{  \lambda (\tilde{\lambda}\tilde{v}^2 + \mu^2 )}{(\lambda v^2 + \mu^2) } \r\}.
\ee
If we now parametrize the fluctuations around the field configurations (\ref{SPP2}) as
\be  \label{SSPtriplettt}
\phi = e^{i \mu t L_3} \l( \begin{array}{c} \phi_0 + \delta \phi^1\\ \delta \phi^2 \\  \delta \phi^3 \end{array} \r), \qquad   \tilde{\phi}  = e^{i \mu t L_3} \l( \begin{array}{c} \delta \tilde{\phi}^1 \\ \tilde{\phi}_0 +\delta \tilde{\phi}^2\\  \delta \tilde{\phi}^3 \end{array} \r),  
\ee
and plug these expression into the Lagrangian (\ref{La2}), we find that the determinant of the inverse propagator matrix has the  form
\be \la{SO(3)2det}
\det [\mbf{D} (\omega, k) ] = (\omega^2 -k^2 -\mu^2)^2 D(\omega, k) \; ,
\ee
where $D(\omega, k)$ is regular for small $\omega$ and $k$.
We immediately see that the spectrum contains two modes with mass $m= \mu$. Incidentally, it is quite remarkable that there are modes that have an exactly relativistic dispersion relation at tree level even though Lorentz symmetry is spontaneously broken by the background (\ref{SPP2}). The function $D(\omega, k)$ on the RHS of equation (\ref{SO(3)2det}) is such that 
\be
\lim_{\omega \to 0} D(\omega, 0) \sim \omega^2 \, v^6, \qquad \qquad \lim_{k \to 0} D(0, k) \sim k^2 \, v^6.
\ee
Based on our discussion in section \ref{sec:spectrum}, we conclude that the spectrum also contains one linear massless mode and three radial modes with mass $m \sim v$. Once again, in the limit $\mu \ll v$ we can integrate out the radial modes and obtain an effective action for the Goldstones which is exactly the one given by the coset construction when no gauge-fixing conditions (other than the boost ones) are imposed. 

The reason why one should not impose any gauge-fixing condition in this case is because of the different mechanism of symmetry breaking.  The order parameter is now given by the pair $(\phi, \tilde{\phi})$ which transforms according to a reducible representation of $SO(3)$. Unlike in the previous example, the low-energy fluctuations of the order parameter are now all independent.  This can be deduced from the fact that no spacetime dependent linear combination of broken generators $f_i(x) \,L_i$ satisfies both
\be
f_i(x) \,L_i \langle \phi \rangle = 0\, ,\,~~ \mbox{and} ~~f_i(x) \,L_i \langle \tilde{\phi} \rangle = 0\, .
\ee
Thus there is no gauge redundancy in the Goldstone parameterization of  the physical fluctuations $\delta \phi_n$ and $\delta \tilde{\phi}_n$.

\subsection{Linear sigma model for (iso)spin-2 tensor}

Finally, let us consider a model in which the $SO(3)$ symmetry is completely broken by a spin-2 representation, i.e.~a symmetric and traceless rank-2 tensor,  acquiring a non-vanishing expectation value. The Lagrangian for this model is:
\be
\mathcal{L} = -\fr{1}{2} \, \d_\mu \Phi^i{}_j \d^\mu \Phi^j{}_i - \lambda \l( \Phi^i{}_j \Phi^j{}_i  - v^2\r)^2 
\ee
We will consider the SSP field configuration
\be \la{SSP3}
\langle \Phi \rangle= e^{i \mu t L_3} \l( \begin{array}{ccc} \Phi_0  & 0& 0 \\ 0 & -\Phi_0  & 0 \\ 0&0&0 \end{array} \r) e^{- i \mu t L_3} .
\ee
with $\Phi_0 = \sqrt{v^2 + \mu^2 /\lambda}$. For simplicity, let us ignore the ``radial" modes and focus directly on the Goldstone modes, by parametrizing fluctuations around the field configuration (\ref{SSP3}) as follows:
\be
\Phi=  e^{i \mu t L_3} e^{i \pi^i L_i} \l( \begin{array}{ccc} \Phi_0 & 0& 0 \\ 0 & -\Phi_0 & 0 \\ 0&0&0 \end{array} \r) e^{- i \pi^i L_i} e^{- i \mu t L_3} .
\ee
The inverse propagator matrix for the  $\pi^i$ modes then is 
\be
\mbf{D} (\omega, k) = \left(
\begin{array}{ccc}
 \omega ^2 - k^2- 3 \mu ^2  & - 2 i \mu \omega  & 0 \\
 2 i \mu \omega & \omega ^2 - k^2- 3 \mu ^2 & 0 \\
 0 & 0 & 4 \left(\omega ^2 - k^2\right)\\
\end{array}
\right)
\ee
We already see that we get one massless Goldstone with linear dispersion relation. By setting to zero the determinant of the upper-left $2\times2$ block we get the dispersion relations for the other two modes:
\be
\omega_\pm = \sqrt{k^2 + 4 \mu^2} \pm \mu.
\ee
Thus, we get two massive modes with masses $m= \mu$ and $m=3 \mu$. It is interesting to see how we always get a linear massless mode and a massive mode with $m =\mu$, as predicted by the coset construction. The second massive mode has a mass $m = 3 \mu$, which is different from the mass we obtained when $SO(3)$ was spontaneously broken by two triplets. This shows explicitly that the mass of the second massive mode depends on the symmetry breaking mechanism, and, more  in general, on the details of the theory.

In this example, the three Goldstone modes are again independent from each other because the equation
\be
f_i(x) \,L_i\langle \Phi \rangle=0
\ee 
cannot be satisfied for any choice of $f_i(x)$ and hence no gauge redundancy exists amongst the Goldstones. Therefore, the low-energy effective action follows from the coset construction without the need to impose any gauge-fixing conditions (other than the boost ones).

\section{Interpolating Fields} \label{inter}

Not only does the standard Goldstone theorem predict the existence of certain excitations, it also gives information about their nature, by associating them with the spontaneously broken currents of the symmetry group, which can serve as the corresponding interpolating fields. In the celebrated  QCD example of $SU(2)_L \times SU(2)_R$ broken  down to the diagonal isospin $SU(2)$, the currents are bilinear in the quark fields, which implies that the related Goldstone particles, the pions, can be thought of as  quark-antiquark  bound states. What are then the interpolating fields for our gapped Goldstones?

\subsection{The ``non-relativistic picture"}

The point of view suggested in~\cite{Nicolis:2011pv,Nicolis:2012vf} and at the basis of our coset construction, is that the ground state $|\mu \rangle$ of a system at finite  density for  a broken charge $Q$, is a state spontaneously breaking both time translations ($H$) and  $Q$, but leaving the combination $H'= H - \mu Q$ unbroken. At the level of the expectation values of relativistic field operators,  $|\mu \rangle$ can be thought of as a field configuration that is spatially homogeneous and evolves in time along a symmetry direction, eq.~\eqref{SSPdef}. With this picture in mind, it seems natural to propose {\it explicitly time dependent} operators as the appropriate interpolating fields of the low energy excitations. In particular, equation~\eqref{SSPdef} suggests to use  operators of the form
\begin{equation}\label{transform}
\bar {\cal O}(x) \ \equiv \ e^{i \mu Q t} \ {\cal O}(x) \ e^{-i \mu Q t}\, ,
\end{equation}
where  ${\cal O}(x)$ is a standard local relativistic operator, in particular, evolving in time with $H$:
\be
\frac{d  {\cal O}(x)}{d t} = i [ H,  {\cal O}(x)] \; .
\ee 
By doing so, not surprisingly, we end up defining quantities that evolve in time with the non relativistic effective Hamiltonian $H'$,
\begin{equation} 
\frac{d \bar {\cal O}(x)}{d t} = i [H', \bar {\cal O}(x)],
\end{equation}
and that are explicitly time-dependent from the point of view of the original relativistic theory.
With this convention for the time evolution of barred operators, which we call ``{non-relativistic (NR) picture}", we can write $n$-point functions
\begin{equation}
\langle \mu | \bar {\cal O}_1(x_1) \bar {\cal O}_2(x_2) \dots \bar {\cal O}_n(x_n)|\mu \rangle\, ,
\end{equation}
with the usual desired properties, such as that of being invariant under a global time translation $t_i \rightarrow t_i + \Delta t$.

What are the NR operators $\bar J^\textsf{a}_\mu(x)$ corresponding to the broken conserved currents $J^\textsf{a}_\mu(x)$? In order to answer this question it is handy to choose directly the basis of broken generators introduced in Sec.~\ref{setup}, distinguishing between commuting and non-commuting broken generators. Commuting generators are untouched by the transformation~\eqref{transform}, 
\be
\bar J_\alpha^\mu(x) =  J_\alpha^\mu(x) \; .
\ee 
On the other hand, for each  pair of non-commuting broken generators we obtain 
\begin{equation} \label{balii}
 \bar J^\mu_{a,l}(x) \ = \ \exp(\mu q_a  t \, i \sigma_2)_{lm} J^\mu_{a,m}(x), \qquad 
 i \sigma_2 =  \left( \begin{array}{cc} 0 & 1 \\
 							-1 & 0 \end{array} \right) \; ,
\end{equation}
with $l,m = \pm$, and $q_a$ is defined in equation~\eqref{block}. Explicitly, the above expression for $\bar J^\mu_{a,\pm}$ is a time-dependent rotation mixing the  $+$ and $-$ components (since there is no mixing between different values of $a$, we drop that index from now on):
\begin{equation}
\begin{split} \label{pedantic} 
\bar J^\mu_{+}(x) \ &=  J^\mu_{+}(x) \cos(\mu q t) + J^\mu_{-}(x) \sin(\mu q t),\\[1mm]
\bar J^\mu_{-}(x) \ &=- J^\mu_{+}(x) \sin(\mu q t) + J^\mu_{-}(x) \cos(\mu q t)\, .
\end{split}
\end{equation}
Due to their explicit time dependence, the operators $\bar J^\mu_{\pm}$ are {\it not} conserved currents. However, they are the appropriate interpolating fields for our gapped Goldstones particles, as we show in Appendix~\ref{B}, 
\begin{equation} \label{vpm}
\langle \mu | \, \bar J^0_{\pm} (t, \vec x) \, |\pi(\vec p) \rangle \ \sim \ v_\pm  \, e^{- i ( E(\vec p) \,  t \, -\,  \vec p \cdot \vec x)} \, ,
\end{equation}
where $E(\vec p)$ is the fixed-gap Goldstone's dispersion relation:  $E(\vec p) = \mu q + {\cal O} (p^2)$.
The NR-currents are defined by~\eqref{balii}-\eqref{pedantic} up to some ``initial conditions"---effectively, some initial time, by the substitution $t\rightarrow t-t_0$. It is not difficult to show that an appropriate choice of $t_0$ sets to zero either of the constants $v_+$, $v_-$. This implies that we can always choose either  $\bar J^\mu_{+}$ or $\bar J^\mu_{-}$ to be the only interpolating field for the fixed-gap Goldstone boson. To be concrete, let's  conventionally  choose $\bar J^\mu_{-}$ as the interpolator. 

As an example, consider the $SO(3)$ single-triplet case of sect.~\ref{sec:so(3)}, with SSP solution~\eqref{SSPtriplettt}. In that example, the gapped Goldstone is clearly identified with the field $\phi_3$ (see also~\cite{Nicolis:2012vf}) and, moreover, we have $q=1$. The $SO(3)$ conserved currents are
\be
J^\mu_i = - \epsilon_{i j k} \, \phi_j \partial^\mu \phi_k \;.
\ee
By identifying $J_1$ with $J_+$ and $J_2$ with $J_-$ we obtain, to first order in perturbations, 
\be
J^\mu_+ \simeq -  \phi_0 \sin (\mu t) \partial^\mu \phi_3 \; , \qquad J^\mu_- \simeq  \phi_0 \cos (\mu t) \partial^\mu \phi_3 \; .
\ee  
Both currents create and annihilate $\phi_3$ quanta, that is, the gapped Goldstone excitations, as predicted.
However, the time dependent combinations~\eqref{pedantic} give a more convenient and less redundant basis:
\be
\bar J^\mu_{+} \simeq 0 \; , \qquad  \bar J^\mu_{-} \simeq  \phi_0   \partial^\mu \phi_3  \; .
\ee

\subsection{Counting the particles in the spectrum}

If the $\mu\rightarrow 0$ limit is smooth, that is, if no phase transitions are encountered,  we expect all our Goldstone excitations---gapless and gapped alike---to become, in that limit, standard massless relativistic Goldstone bosons. The latter are as many as the broken generators at $\mu =0$. So, if we now go the other way, when we turn on the chemical potential $\mu$, the number of broken generators in general increases,  but the number of Goldstone excitations remains constant.

By looking again at the $SO(3)$ single-triplet example we notice that, at $\mu=0$, $J_2$ and $J_3$ are broken, while $J_1$ is unbroken. By turning on $\mu$ along the $J_3$ direction, we break $SO(3)$ completely. However, the NR-current associated with $J_1$, ($\bar J_+$ in the discussion above) still does not interpolate any particle: the total number of light degrees of freedom is conserved. 
Vice-versa, in those examples where $SO(3)$ is broken completely already at $\mu = 0$---the two-triplet or isospin-2 case---$\bar J_+$ is also interpolating a particle, although different from the fixed-gap Goldstone $|\pi(\vec p)\rangle$: the  coefficient $v_+$ in eq.~\eqref{vpm}  still vanishes. It  interpolates an {\it unfixed-gap} Goldstone, the partner of the fixed-gap one interpolated by $\bar	 J_-$.

Fixing the gauge the $D_0 \pi^a_-=0$ at the level of the coset construction (Sec.~\ref{sec:pipm}), is equivalent to stating that $\bar J_+$ does not interpolate any particle. Although from the point of view of the coset construction at finite $\mu$ we seem to be completely free to treat $\pi^a_+$ as redundant or physical, the present discussion suggests---in the cases in which we know the $\mu \to 0$ limit to be smooth---to look at the number of broken generators at $\mu =0$ first, and choose  covariant unitary gauge for all the Goldsone fields associated with the generators that {\it become} broken when $\mu$ is turned on.

More generally, we can relate these arguments to those of the inverse Higgs/gauge redundancy in the following way.  If a current of broken generators does not interpolate a Goldstone (as is the case for $\bar J_+$ in the $SO(3)$ single-triplet example above), then we expect
\be
\bar J_+ \langle \Phi \rangle = f_i(x) X_i \langle \Phi \rangle = 0 \, ,
\ee
where $\Phi$ is the order parameter and $f_i(x) X_i$ is simply $\bar J_+ $ written in a basis of broken generators $X_i$.  This immediately implies a gauge symmetry for the fluctuation
\be
\delta \Phi \simeq i \pi_i X_i \langle \Phi \rangle \, ,
\ee
of the form
\be
\pi_i \rightarrow \pi_i +\Lambda(x)  f_i(x) \, .
\ee
This gauge symmetry is responsible for removing the spurious Goldstone bosons, giving the correct overall counting of degrees of freedom.


\section{Energy Considerations}

The appearance of gapped Goldstones is a direct consequence of having shifted the Hamiltonian $H \rightarrow H'=H- \mu Q$.  As mentioned above, such a procedure is necessary when $Q$ and thus $H$ are broken: in that case, excitations can only be classified in terms of their ``energies" as measured by the unbroken combination $H'$.  But it is natural to ask what happens when {\it unbroken} charges are at finite density, say some $T$.  The ground state of this system still corresponds to the lowest eigenvalue of the operator $H' = H- \mu T$.  Yet because $T$ and $H$ are unbroken, the eigenstates of $H'$ can  be chosen to be also eigenstates of the original Hamiltonian $H$, as well as of $T$. Then, if one uses $H$ to classify excitations, one can run the theorem of \cite{Nicolis:2012vf} again and discover that all  the once fixed-gap Goldstones are now gapless, in agreement with more standard Goldstone theorems. In this case there seems to be an ambiguity in how we define ``energy"---should we use $H$ or $H'$?  Whether or not Goldstones are gapped would depend on which operator one uses. Yet, ultimately, any  prediction for the outcome  of an actual experiment should be independent of such a choice. 

In this section we address this issue. Before we look in detail at some aspects of it, it is worth pointing out that which definition of energy is the natural one to use, sometimes just depends on the question one asks.
For instance, for thermodynamical considerations, since the combination $H -\mu T$ is precisely what appears in the partition function, $H'$ probably provides the more convenient definition of energy for the excitations, even though one should keep in mind that in thermodynamical relations like $E + PV = \mu N + T S$, $E$ always stands for the expectation value of the original Hamiltonian $H$. 
Another convenient feature of $H'$ is that it is minimized by our state $| \mu \rangle$, so that all excitations are positive energy according to $H'$, but not necessarily according to $H$.
On the other hand, if one is interested in how gravity couples to our excitations, for instance, if one wants to consider cosmological applications of our Goldstone system, then $H$ is probably the more convenient measure of energy, since gravity couples to the stress-energy tensor $T^{\mu\nu}$ and, as we will see below, $H$ is nothing but the spatial integral of $T^{00}$ (see also a related discussion in \cite{ArkaniHamed:2003uy}).

\subsection{Probe-ability}\label{prob}
There is one aspect of being ``gapped" that seems to be very concrete, and not just a matter of definition: if, according to some definition of energy, say $H'$, an excitation is gapped, then that excitation cannot be produced by working below the gap---one cannot probe it directly at low energies. On the other hand, if one now changes one's definition of energy and uses $H$, the excitation in question becomes gapless, and now it {\it can} be probed at arbitrarily low energies. 

Notice that the resolution of the apparent paradox is not simply that what we mean by ``low energies'' is different in the two cases: we might be using low-energy probes, like for instance external photons, that are neutral under $T$, and thus completely insensitive to the change of Hamiltonian. If we stick to this case, this must mean that in the second picture, the excitation cannot be probed because of {\it another} conservation law, i.e., not energy conservation, but {\it charge} conservation. The charge in question is $T$ itself of course: by assumption, the external probe carries no $T$, while the Goldstone excitation under consideration carries the same charge $q$ as the broken generator  it is associated with. But recall that the same $q$ also determines the gap in the $H'$ picture:
\be
T | \pi \rangle = q | \pi \rangle  \; , \qquad H'  | \pi \rangle = \mu q  | \pi \rangle \; , \qquad H | \pi \rangle = 0
\ee
(we are implicitly subtracting the charge and energies of the ground state $|\mu \rangle$.)
And so, all Goldstones that are fixed-gap in the $H'$ picture, carry positive charge under $T$ (when $\mu$ is positive), and cannot be produced {\it in any number}, and at {\it any energy} in processes like
\be
\gamma \gamma \to \pi_{\rm fg} \pi_{\rm fg} \dots \pi_{\rm fg} \; ,
\ee
where $\gamma$ stands for an external neutral probe particle---e.g.~a photon---and $\pi_{\rm fg}$ for a fixed-gap Goldstone. In other words, their production is forbidden because of charge conservation, regardless of the energies involved.

Are these excitations completely unprobe-able from the outside? If so, why are we talking about them?
Fortunately, their unfixed-gap partners save the day. Recall that for each fixed-gap Goldstone, there is a potential unfixed-gap partner (see sect.~\ref{sec:pipm}). In general this can be a redundant degree of freedom, removable by fixing a gauge, but {\it not} when $T$ is unbroken: in such a case, the Goldstone fields $\pi_a^\pm$ transform linearly as a doublet under $T$, and it inconsistent with the unbroken $T$ to remove one and not the other\footnote{Alternatively, one could take a complex linear combination $\phi_a = \pi_a^+ + i \pi_a^-$, it terms of which the candidate gauge-fixing condition would take the form
\be
0 = D_0 \phi_a \simeq \partial_t \phi_a + i \mu q_a \phi_a \; ,
\ee
which---if imposed---would completely determine the time-dependence of the full complex field $\phi_a$, thus effectively eliminating {\it two} degrees of freedom.}. Now, the crucial property of these unfixed-gap partners $\pi_{\rm ug}$, for our discussion, is that they carry a charge under $T$ that is exactly opposite to that carried by the associated fixed-gap excitations. As a result, pair-production processes like
\be
\gamma \gamma \to \pi_{\rm fg} \pi_{\rm ug}
\ee
are allowed, both by charge conservation, and, at high enough energies, by energy conservation as well.
Notice that, since the energies of the external $\gamma$'s are insensitive to which Hamiltonian we are using, the energy threshold for the process to happen has to be insensitive as well. It is, since
\be \label{threshold}
E'(\pi_{\rm fg}) + E'(\pi_{\rm ug}) = E(\pi_{\rm fg}) + E(\pi_{\rm ug}) \; ,
\ee
where we used that the two Goldstone excitations carry opposite $T$-charges.

The fact that the pair-production energy threshold is the same in the two pictures, can be used to bound the gap of $\pi_{\rm ug}$ in the $H$-picture, which is that used by Nielsen and Chadha for instance \cite{Nielsen:1975hm}. The LHS of \eqref{threshold} is always bigger than $E'(\pi_{\rm fg}) = \mu q$, because $H'$ is minimized by our finite-density state $| \mu \rangle$, and so $E'(\pi_{\rm ug})$ has to be positive. On the RHS, the first term vanishes, because all fixed-gap Goldstones are gapless in the  $H$-picture. This means that
\be
E(\pi_{\rm ug}) > \mu q \; ,
\ee
which applies to all unfixed-gap partners of fixed-gap Goldstones. Once again, this result is non-perturbatively {\it exact}.

\subsection{Gravitational energy}
Often, what we mean by ``energy" is ultimately gravitational energy, i.e., the $00$-component of the gravitational stress-energy tensor $T_{\mu\nu}^G$.  However, for systems at finite density, the gravitational stress-energy tensor does not necessarily coincide with the canonical stress-energy tensor $T_{\mu\nu}^c$ one derives via Noether's theorem.

To see this, consider a complex scalar field at finite chemical potential:
\be
\label{Lagr}
{\cal L} = |D_\mu \Phi|^2 -V \big( |\Phi|^2 \big) \, ,
\ee
where $D_\mu = \partial_\mu+i\mu \, \delta_\mu^0$.  Canonically conjugate momenta are given by 
\be
\Pi \equiv \frac{\delta {\cal L}}{\delta{\dot{\Phi}}} = (D_0 \Phi)^* \, , ~~~~\Pi^* \equiv \frac{\delta {\cal L}}{\delta{\dot{\Phi}}^*} = D_0 \Phi \, .
\ee
The conserved current is given by $J^0 =i( \Pi^* \Phi^*-\Pi \Phi)$.  The Hamiltonian density ${\cal H}' = {\cal H}-\mu J_0$ coincides with the $00$-component of the ``canonical" stress-energy tensor:
\be
T_{\mu\nu}^c \equiv \frac{\delta {\cal L}}{\delta\,\partial_\mu \psi^a}\partial_\nu \psi^a-g_{\mu\nu}{\cal L} \,.
\ee
Using the Lagrangian given in (\ref{Lagr}), one has
\be
T_{00}^c =|\Pi|^2 +|D_j \Phi |^2+V \big( |\Phi|^2 \big) - i \mu (\Pi^* \Phi^*-\Pi \Phi) = {\cal H}-\mu J^0\, ,
\ee
as we expect.

At finite chemical potential, the Hamiltonian ${\cal H}'$ does not coincide with the $00$-component of the gravitational stress-energy tensor, defined as
\be
T_{\mu\nu}^G \equiv \frac{2}{\sqrt{-g}}\frac{\delta(\sqrt{-g} {\cal L})}{\delta\,g^{\mu\nu}} \; .
\ee
Instead, it is ${\cal H}$ that coincides with the gravitational stress-energy tensor. Indeed, using the Lagrangian (\ref{Lagr}), and coupling our scalar to a generic metric via $|D_\mu \Phi|^2 \to g^{\mu\nu} D_\mu \Phi  D_\nu \Phi^*$,  one finds
\be
T_{00}^G = |\Pi|^2 +|D_j \Phi |^2+V \big(|\Phi |^2\big)= {\cal H} \,.
\ee
Thus for gravitational considerations, $H$ is perhaps the more relevant measure of energy.

\section{Outlook}
We would like to conclude our paper by emphasizing the generality of our results: they apply to any relativistic theory that exhibits spontaneous symmetry breaking in the presence of a finite density for one of the broken charges.  As such, they have potential applications to systems as diverse as non-abelian superfluid systems like QCD at finite isospin density \cite{Son:2000xc} and inflationary cosmology with  internal non-abelian symmetries \cite{Nicolis:2011pv}.  We plan to investigate these applications in the near future.

The previous work of ref.~\cite{Watanabe:2013uya} extends the results on linear-gapless, quadratic-gapless, and fixed-gap Goldstones to non-relativistic systems. However, as we emphasized in the introduction, in the real world Poincar\'e invariance is broken always spontaneously, and so we feel that a complete analysis of realistic non-relativistic systems should take this into account. For instance, it was our emphasis on the spacetime symmetries that made us discover the fourth class of Goldstones---the unfixed-gap ones---which do not appear in the analysis of \cite{Watanabe:2013uya}.\footnote{These Goldstones are now mentioned in a footnote in a revised version of the manuscript~\cite{Watanabe:2013uya}.} However, in our analysis we assumed that Poincar\'e was broken only by the presence of a finite charge density. To extend our analysis to standard  non-relativistic systems in the lab, we should include in our coset construction the degrees of freedom associated with an independent breaking of Poincar\'e invariance, say the phonons of an underlying medium, and see whether and to what extent these modify our results. 

Finally, while the interpretation of the inverse Higgs constraints presented here is complementary to those in the previous literature \cite{Ivanov:1975zq,Low:2001bw,McArthur:2010zm}, it would be useful to formalize this correspondence.  In particular, one would like to derive the non-linear form of the Goldstone gauge redundancies  from the symmetry algebra.  It would also be interesting to see whether it is possible (and useful) to write down gauge-invariant actions for the Goldstone fields that are also invariant under all unbroken global symmetries.
 These issues are  the subject of future work.

\section*{Acknowledgments}  
We would like to thank Garrett Goon, Austin Joyce, Hitoshi Murayama, Francesco Nitti and Haruki Watanabe for useful discussions. We are especially thankful to Kurt Hinterbichler for collaboration in the early stages of this project. Finally, we are very grateful to Daniel Green and Ira Rothstein for bringing to our attention refs.~\cite{Alday:2010ku,Kapustin:2012cr}, and to Tomas Brauner for carefully pointing out the numerous typos contained in the first version of this manuscript. FP thanks the Department of Physics at Columbia University for hospitality. The work of AN, RP, and RAR is supported by NASA under contract NNX10AH14G and by the DOE under contract DE-FG02-11ER41743.

\appendix

\section{Constraints from Jacobi Identity}\label{app1}

The structure constants $f_{IJK}$ are not completely arbitrary, but obey some constraints that follow from the Jacobi identity
\be \la{jacobi}
\mathcal{J}_{IJLM} \equiv f_{IJK} f_{LKM} + f_{LIK} f_{JKM} + f_{JLK} f_{IKM} = 0.
\ee
In particular, we will find the following results useful:
\begin{enumerate}
\item The first result is about structure constants involving two commuting and one non-commuting broken generators: 
\be
\mu_X \mathcal{J}_{\beta,\gamma, X, a_\mp} + \mu_T \mathcal{J}_{\beta,\gamma, T, a_\mp} = 0 \qquad \Longrightarrow \qquad   f_{\beta\gamma a_\pm} = 0.
\ee
\item The second result can be derived by replacing $\beta$ with $X$ in the previous equation: 
\be
\mu_X \mathcal{J}_{X,\gamma, X, a_\mp} + \mu_T \mathcal{J}_{X,\gamma, T, a_\mp} = 0 \qquad \Longrightarrow \qquad   f_{X\gamma a_\pm} = 0.
\ee
\item The third result is about structure constants involving one commuting and two non-commuting broken generators: 
\be
\mu_X \mathcal{J}_{\alpha,a_+, X, b_-} + \mu_T  \mathcal{J}_{\alpha,a_+, T, b_-} = 0 \qquad \Longrightarrow \qquad   f_{\alpha a_+ b_+} q_b = f_{\alpha a_- b_-} q_a.
\ee
Thus, we either have $f_{\alpha a_+ b_+} =f_{\alpha a_- b_-} =0$, or we can divide this equation by the same equation with $a \leftrightarrow b$ and, using the antisymmetry of the structure constants we get $q_a^2 = q_b^2$. Since $q_a >0$ by construction, this implies $q_a = q_b$ and therefore $f_{\alpha a_+ b_+} = f_{\alpha a_- b_-}$. Thus, we conclude that 
\begin{subequations}
\ba
&f_{\alpha a_+ b_+} = f_{\alpha a_- b_-}, &\\
&f_{\alpha a_+ b_+} \neq 0 \quad \Longrightarrow \quad q_a = q_b.&
\ea
\end{subequations}
\item The result above can be easily rederived with $\alpha$ replaced by $X$:
\begin{subequations}
\ba
&f_{X a_+ b_+} = f_{X a_- b_-}, &\\
&f_{X a_+ b_+} \neq 0 \quad \Longrightarrow \quad q_a = q_b.&
\ea
\end{subequations}
\item The next result is again about structure constants involving one commuting and two non-commuting broken generators: 
\begin{subequations}
\ba
\mu_X \mathcal{J}_{\alpha,a_+, X, b_+} + \mu_T  \mathcal{J}_{\alpha,a_+, T, b_+} = 0 \quad &\Longrightarrow& \quad   f_{\alpha a_+ b_-} q_b = - f_{\alpha a_- b_+} q_a \\
\mu_X \mathcal{J}_{\alpha,a_-, X, b_-} + \mu_T  \mathcal{J}_{\alpha,a_-, T, b_-} = 0 \quad &\Longrightarrow& \quad   f_{\alpha a_- b_+} q_b = - f_{\alpha a_+ b_-} q_a. \la{eq}
\ea
\end{subequations}
By subtracting these two equations we get
\be
f_{\alpha a_- b_+} (q_b -q_a)= f_{\alpha a_+ b_-} (q_b -q_a),
\ee
Thus, if $q_a \neq q_b$ we obtain
\be
f_{\alpha a_- b_+} =  f_{\alpha a_+ b_-} \qquad \mbox{for} \qquad q_a \neq q_b
\ee
If we now use this result in equation (\ref{eq}) and we remember that all $q_a$'s are positive by construction, we obtain 
\be
f_{\alpha a_+ b_-} = 0 \qquad \mbox{for} \qquad q_a \neq q_b.
\ee
If instead $q_a = q_b$, we get from (\ref{eq}) that
\be
f_{\alpha a_- b_+} =  - f_{\alpha a_+ b_-} \qquad \mbox{for} \qquad q_a = q_b.
\ee
Thus, we conclude that 
\begin{subequations}
\ba
&f_{\alpha a_- b_+} = f_{\alpha b_- a_+}, &\\
&f_{\alpha a_- b_+} \neq 0 \quad \Longrightarrow \quad q_a = q_b.&
\ea
\end{subequations}
\item The result above can be easily rederived with $\alpha$ replaced by $X$:
\begin{subequations}
\ba
&f_{X a_- b_+} = f_{X b_- a_+}, &\\
&f_{X a_- b_+} \neq 0 \quad \Longrightarrow \quad q_a = q_b.&
\ea
\end{subequations}
\end{enumerate}

\section{Broken Current Matrix Elements}
\label{B}
In this appendix we show that the NR currents defined in Sec.~\ref{inter} interpolate the gapped Goldstone states. 

The most general proof of the Goldstone theorem develops from the constancy in time of matrix elements of the form 
\be
\kappa_{I} \equiv \langle \mu |[Q_I(t), A (0)] | \mu \rangle \, ,
\ee
for some local operator $A(x)$ and where $Q_I$ is a broken conserved charge.
The constancy in time of $\kappa_I$ is guaranteed by current conservation and by the relativistic structure of the theory. Its being non zero for some order parameter $A(x)$ is the statement that the $Q_I$ is spontaneously broken by the state $| \mu \rangle$. In the case of broken, non-commuting charges $Q_{a, \pm}$, the possible time dependence of the corresponding $\kappa_{a, \pm}$ can be distilled into the  expression
\be \label{a_2}
\kappa_{a, \pm} = \sum_N e^{- i  E_N (\vec p = 0)\,  t}  \ \langle \mu |\,  e^{i \mu Q t} \, J^0_{a,\pm} (0) \, e^{-i \mu Q t} \,  |N,\vec p = 0\rangle \langle N,\vec p = 0 | \, A(0)\, |\mu \rangle -  {\rm c.c.} \, ,
\ee
where the sum is over intermediate momentum eigenstates. Such eigenstates are chosen to be also eigenstates of $H'$, with eigenvalues (dispersion relations) $E_N(|\vec p|)$.
From~\eqref{a_2}, after straightforward manipulations, one can show~\cite{Nicolis:2012vf} that in order for $\kappa_{a, \pm}$ to be constant {\it and} different from zero, there must exist a state $|\pi(\vec p)\rangle$ in the theory with $E_N (\vec p = 0) = \mu q_a$. This is the fixed-gap Goldstone.

We can now  focus directly on 
such a state.
From~\eqref{a_2}, it  follows that
\begin{equation} \label{a_3}
C_{a, \pm} \ =\  (2 \pi)^3 e^{- i  E ( 0)\,  t}  \ \langle \mu |\,  e^{i \mu Q t} \, J^0_{a,\pm} (0) \, e^{-i \mu Q t} \,  |\pi(\vec p=0) \rangle 
\end{equation}
is a time-independent complex number different from zero. Now we want to express $C_{a, \pm}$ in terms of matrix elements of the ``tilded" currents defined in Sec.~\ref{inter}. Since they evolve with $H'$, we have $J^0_{a,\pm} (0) = e^{- i H' t} \bar J^0_{a,\pm} (t, \vec 0) e^{ i H' t} $. After this substitution, the strategy is in a sense to ``rewind"  the derivation in~\cite{Nicolis:2012vf}, and reintroduce the spatial dependence in $\bar J$. This is done by expressing the $\vec p =0$ condition in~\eqref{a_3} by integrating over a delta function $\delta^3(\vec p)  = (2 \pi)^{-3} \int d^3 x e^{i \vec p \cdot \vec x}$, 
\begin{equation}
C_{a, \pm} \ =\  \int \! d^3 p \, d^3 x  \, e^{- i  E \,  t}  e^{ i  \vec p \cdot \vec x} \ \langle \mu |\,  e^{i \mu Q t} \, \bar J^0_{a,\pm} (t, \vec 0) \, e^{-i \mu Q t} \, e^{ i H' t}  |\pi(\vec p) \rangle \, ,
\end{equation}
where we used that $H'$ commutes with $Q$, and that it annihilates $| \mu \rangle$.
The exponentials  outside the matrix element can be transformed into the corresponding operators hitting the state $|\pi(\vec p) \rangle$ inside the matrix element. The energy $E$ cancels with $H'$ and the exponential of the momentum operator can be used to translate $\bar J^0_{a,\pm} (t, \vec 0)$ at the point $\vec x$---using once again that the momentum commutes with $Q$ and annihilates $| \mu \rangle$.
On the other hand, $\mu Q$ hitting $\langle \mu |$ produces $H$:
\begin{equation}
C_{a, \pm} \ =\  \int \! d^3 p \, d^3 x  \,  \ \langle \mu |\,  e^{i H t} \, \bar J^0_{a,\pm} (t, \vec x) \, e^{-i \mu Q t}  |\pi(\vec p) \rangle \, .
\end{equation}
The integral of $\bar J^0_{a,\pm} (t, \vec x)$ is a time dependent combination of conserved charges as follows from eq.~\eqref{pedantic}, and therefore commutes with $H$. 
Hence we finally obtain
\begin{equation}
C_{a, \pm} \ =\  \int \! d^3 p \, d^3 x  \  e^{i  E(|\vec p|) \,  t} \ \langle \mu | \, \bar J^0_{a,\pm} (t, \vec x) \, |\pi(\vec p) \rangle \, .
\end{equation}
Because translations are not broken, the $\vec x$ dependence of the above matrix element is simply
 $e^{ i  \vec p \cdot \vec x}$.  
We deduce that $\bar J$ interpolates the state $|\pi(\vec p) \rangle$ is the usual sense:
\begin{equation}
\langle \mu | \, \bar J^\mu_{a,\pm} (t, \vec x) \, |\pi(\vec p) \rangle \ \sim  e^{- i ( E \,  t \, -\,  \vec p \cdot \vec x)} \, .
\end{equation}

\bibliographystyle{utphys}
\addcontentsline{toc}{section}{References}
\bibliography{gaps}

\providecommand{\href}[2]{#2}\begingroup\raggedright\begin{thebibliography}{10}

\bibitem{Nicolis:2012vf}
A.~Nicolis and F.~Piazza, ``{A relativistic non-relativistic Goldstone theorem:
  gapped Goldstones at finite charge density},''
  \href{http://dx.doi.org/10.1103/PhysRevLett.110.011602,
  10.1103/PhysRevLett.110.039901}{{\em Phys.Rev.Lett.} {\bfseries 110} (2013)
  011602},
\href{http://arxiv.org/abs/1204.1570}{{\ttfamily arXiv:1204.1570 [hep-th]}}.

\bibitem{Callan:1969sn}
J.~Callan, Curtis~G., S.~R. Coleman, J.~Wess, and B.~Zumino, ``{Structure of
  phenomenological Lagrangians. 2.},''
\href{http://dx.doi.org/10.1103/PhysRev.177.2247}{{\em Phys.Rev.} {\bfseries
  177} (1969) 2247--2250}.

\bibitem{Coleman:1969sm}
S.~R. Coleman, J.~Wess, and B.~Zumino, ``{Structure of phenomenological
  Lagrangians. 1.},''
\href{http://dx.doi.org/10.1103/PhysRev.177.2239}{{\em Phys.Rev.} {\bfseries
  177} (1969) 2239--2247}.

\bibitem{Volkov:1973vd}
D.~V. Volkov, ``{Phenomenological Lagrangians},''
{\em Fiz.Elem.Chast.Atom.Yadra} {\bfseries 4} (1973) 3--41.

\bibitem{Ogievetsky:1974ab}
V.~I. Ogievetsky, ``{Nonlinear realizations of internal and space-time
  symmetries},'' in {\em {Proceedings of the X-th winter school of theoretical
  physics in Karpacz}}, vol.~1, p.~227.
\newblock Universitas Wratislaviensis, Wroclaw, 1974.

\bibitem{Nielsen:1975hm}
H.~B. Nielsen and S.~Chadha, ``{On How to Count Goldstone Bosons},''
\href{http://dx.doi.org/10.1016/0550-3213(76)90025-0}{{\em Nucl.Phys.}
  {\bfseries B105} (1976) 445}.

\bibitem{Watanabe:2011ec}
H.~Watanabe and T.~Brauner, ``{On the number of Nambu-Goldstone bosons and its
  relation to charge densities},''
  \href{http://dx.doi.org/10.1103/PhysRevD.84.125013}{{\em Phys.Rev.}
  {\bfseries D84} (2011) 125013},
\href{http://arxiv.org/abs/1109.6327}{{\ttfamily arXiv:1109.6327 [hep-ph]}}.

\bibitem{Watanabe:2012hr}
H.~Watanabe and H.~Murayama, ``{Unified Description of Nambu-Goldstone Bosons
  without Lorentz Invariance},''
  \href{http://dx.doi.org/10.1103/PhysRevLett.108.251602}{{\em Phys.Rev.Lett.}
  {\bfseries 108} (2012) 251602},
\href{http://arxiv.org/abs/1203.0609}{{\ttfamily arXiv:1203.0609 [hep-th]}}.

\bibitem{Hidaka:2012ym}
Y.~Hidaka, ``{Counting rule for Nambu-Goldstone modes in nonrelativistic
  systems},'' \href{http://dx.doi.org/10.1103/PhysRevLett.110.091601}{{\em
  Phys.Rev.Lett.} {\bfseries 110} (2013) 091601},
\href{http://arxiv.org/abs/1203.1494}{{\ttfamily arXiv:1203.1494 [hep-th]}}.

\bibitem{Schafer:2001bq}
T.~Sch{\"a}fer, D.~Son, M.~A. Stephanov, D.~Toublan, and J.~Verbaarschot,
  ``{Kaon condensation and Goldstone's theorem},''
  \href{http://dx.doi.org/10.1016/S0370-2693(01)01265-5}{{\em Phys.Lett.}
  {\bfseries B522} (2001) 67--75},
\href{http://arxiv.org/abs/hep-ph/0108210}{{\ttfamily arXiv:hep-ph/0108210
  [hep-ph]}}.

\bibitem{Alday:2010ku}
L.~F. Alday, D.~Gaiotto, J.~Maldacena, A.~Sever, and P.~Vieira, ``{An Operator
  Product Expansion for Polygonal null Wilson Loops},''
  \href{http://dx.doi.org/10.1007/JHEP04(2011)088}{{\em JHEP} {\bfseries 1104}
  (2011) 088},
\href{http://arxiv.org/abs/1006.2788}{{\ttfamily arXiv:1006.2788 [hep-th]}}.

\bibitem{Ivanov:1975zq}
E.~Ivanov and V.~Ogievetsky, ``{The Inverse Higgs Phenomenon in Nonlinear
  Realizations},''
{\em Teor.Mat.Fiz.} {\bfseries 25} (1975) 164--177.

\bibitem{Low:2001bw}
I.~Low and A.~V. Manohar, ``{Spontaneously broken space-time symmetries and
  Goldstone's theorem},''
  \href{http://dx.doi.org/10.1103/PhysRevLett.88.101602}{{\em Phys.Rev.Lett.}
  {\bfseries 88} (2002) 101602},
\href{http://arxiv.org/abs/hep-th/0110285}{{\ttfamily arXiv:hep-th/0110285
  [hep-th]}}.

\bibitem{Watanabe:2013uya}
H.~Watanabe, T.~Brauner, and H.~Murayama, ``{Massive Nambu-Goldstone Bosons},''
\href{http://arxiv.org/abs/1303.1527}{{\ttfamily arXiv:1303.1527 [hep-th]}}.

\bibitem{Kapustin:2012cr}
A.~Kapustin, ``{Remarks on nonrelativistic Goldstone bosons},''
\href{http://arxiv.org/abs/1207.0457}{{\ttfamily arXiv:1207.0457 [hep-ph]}}.

\bibitem{Nicolis:2011pv}
A.~Nicolis and F.~Piazza, ``{Spontaneous Symmetry Probing},''
  \href{http://dx.doi.org/10.1007/JHEP06(2012)025}{{\em JHEP} {\bfseries 1206}
  (2012) 025},
\href{http://arxiv.org/abs/1112.5174}{{\ttfamily arXiv:1112.5174 [hep-th]}}.

\bibitem{framids}
A.~Nicolis, R.~Penco, F.~Piazza, R.~Rattazzi, and R.~Rosen. {\it In
  preparation}.

\bibitem{McArthur:2010zm}
I.~McArthur, ``{Nonlinear realizations of symmetries and unphysical Goldstone
  bosons},'' \href{http://dx.doi.org/10.1007/JHEP11(2010)140}{{\em JHEP}
  {\bfseries 1011} (2010) 140},
\href{http://arxiv.org/abs/1009.3696}{{\ttfamily arXiv:1009.3696 [hep-th]}}.

\bibitem{son}
D.~Son, ``{Low-energy quantum effective action for relativistic superfluids},''
\href{http://arxiv.org/abs/hep-ph/0204199}{{\ttfamily arXiv:hep-ph/0204199
  [hep-ph]}}.

\bibitem{det}
P.~Powell, ``{Calculating Determinants of Block Matrices},''
  \href{http://arxiv.org/abs/1112.4379}{{\ttfamily arXiv:1112.4379 [math.RA]}}.

\bibitem{Miransky:2001tw}
V.~Miransky and I.~Shovkovy, ``{Spontaneous symmetry breaking with abnormal
  number of Nambu-Goldstone bosons and kaon condensate},''
  \href{http://dx.doi.org/10.1103/PhysRevLett.88.111601}{{\em Phys.Rev.Lett.}
  {\bfseries 88} (2002) 111601},
\href{http://arxiv.org/abs/hep-ph/0108178}{{\ttfamily arXiv:hep-ph/0108178
  [hep-ph]}}.

\bibitem{ArkaniHamed:2003uy}
N.~Arkani-Hamed, H.-C. Cheng, M.~A. Luty, and S.~Mukohyama, ``{Ghost
  condensation and a consistent infrared modification of gravity},''
  \href{http://dx.doi.org/10.1088/1126-6708/2004/05/074}{{\em JHEP} {\bfseries
  0405} (2004) 074},
\href{http://arxiv.org/abs/hep-th/0312099}{{\ttfamily arXiv:hep-th/0312099
  [hep-th]}}.

\bibitem{Son:2000xc}
D.~Son and M.~A. Stephanov, ``{QCD at finite isospin density},''
  \href{http://dx.doi.org/10.1103/PhysRevLett.86.592}{{\em Phys.Rev.Lett.}
  {\bfseries 86} (2001) 592--595},
\href{http://arxiv.org/abs/hep-ph/0005225}{{\ttfamily arXiv:hep-ph/0005225
  [hep-ph]}}.

\end{thebibliography}\endgroup

\end{document}